\documentclass[aps,pre,twocolumn,groupedaddress]{revtex4-1}

\usepackage{amsfonts,amssymb,amsmath} 
\usepackage{color} 
\usepackage{graphicx} 
\usepackage{epstopdf,mathrsfs} 
\usepackage[colorlinks,linktocpage,linkcolor=blue]{hyperref}

\begin{document}


\title{Subdiffusion in an external force field}

\author{Yao Chen}
\author{Xudong Wang}
\author{Weihua Deng}

\affiliation{School of Mathematics and Statistics, Gansu Key Laboratory
of Applied Mathematics and Complex Systems, Lanzhou University, Lanzhou 730000,
P.R. China}

\begin{abstract}
The phenomena of subdiffusion are widely observed in physical and biological systems. To investigate the effects of external potentials, say, harmonic potential, linear potential, and time dependent force, we study the subdiffusion described by subordinated Langevin equation with white Gaussian noise, or equivalently, by the single Langevin equation with compound noise. If the force acts on the subordinated process, it keeps working all the time; otherwise, the force just exerts an influence on the system at the moments of jump. Some common statistical quantities, such as, the ensemble and time averaged mean squared displacement, position autocorrelation function, correlation coefficient, generalized Einstein relation, are discussed to distinguish the effects of various forces and different patterns of acting. The corresponding Fokker-Planck equations are also presented. All the stochastic processes discussed here are non-stationary, non-ergodicity, and aging.

\end{abstract}


\maketitle

\section{Introduction}
In the natural world, it is hard to find the real free particles; actually almost all the time, they are in some kinds of external potentials. The motion of particles in complex disordered systems generally is no longer Brownian, exhibiting anomalous diffusion behavior \cite{JeonMetzler:2012,EuleFriedrich:2009,CairoliBaule:2015,MagdziarzWeronKlafter:2008,ChenWangDeng:2017,FedotovKorabel:2015}, which is characterized by the nonlinear evolution in time of the mean squared displacement (MSD) of particles; i.e.,
\begin{equation}\label{1}
\langle(\Delta y(t))^2\rangle=\langle[y(t)-\langle y(t)\rangle]^2\rangle\simeq t^\beta  \quad  (\beta\neq1),
\end{equation}
which represents subdiffusion for $0<\beta<1$ and superdiffusion for $\beta>1$; for the case $\beta=2$, it is called ballistic diffusion and $\beta=0$ the localization diffusion \cite{MetzlerKlafter:2000,BouchaudGeorges:1990}.

One of the most powerful and popular models to describe anomalous diffusion is {continuous-time} random walk (CTRW), which was originally introduced by Montroll and Weiss in 1965 \cite{MontrollWeiss:1965}, extending regular random walks on lattices to a {continuous-time} variable. It has been successfully applied in various fields, such as, the charge carrier transport in amorphous semiconductors \cite{ScherMontroll:1975}, electron transfer \cite{Nelson:1999}, dispersion in turbulent systems \cite{SolomonWeeksSwinney:1993}, and so on.

Another special model to describe the complex dynamics is Langevin equation; its classical version should be the differential equation form of Newton's second law. Compared with CTRW, the Langevin picture has a striking advantage in characterizing external fields. Of course, it also has a close connection with CTRW model. In 1994, Fogedby \cite{Fogedby:1994} used the stochastic time changed method to introduce an equivalent form of the continuum limit of the subdiffusive CTRW --- a Langevin equation coupled with a subordinator, i.e.,
\begin{equation}\label{free1}
\dot{x}(s)=\sqrt{2\sigma}\xi(s), \qquad \dot{t}(s)=\eta(s),
\end{equation}
where $x(s)$ is named as original process with respect to internal time $s$, $\xi(s)$ is a white Gaussian noise with null mean value and autocorrelation function $\langle \xi(s)\xi(s')\rangle=\delta(s-s')$, and $\eta(s)$ is a fully skewed $\alpha$-stable L\'{e}vy noise with $0<\alpha<1$ \cite{SchertzerLarchevequeDuanYanovskyLovejoy:2001} and usually regarded as the formal derivative of the $\alpha$-stable subordinator $t(s)$ \cite{Applebaum:2009}. The time changed process $y(t):=x(s(t))$, where $s(t)$ is the inverse $\alpha$-stable subordinator \cite{KumarVellaisamy:2015,AlrawashdehKellyMeerschaertScheffler:2017}, is an equivalent stochastic trajectory to the continuum limit of the CTRW with power-law distributed waiting times.
Since then, {the subordination} \cite{Applebaum:2009}, which was put forward by Bochner \cite{Bochner:1949} in 1949, has become a useful method to describe the time-changed stochastic processes exhibiting anomalous diffusion. Especially in recent years, the coupled Langevin equations have been widely investigated \cite{MeerschaertScheffler:2004,GajdaWylomanska:2015,WylomanskaKumarPoloczanskiVellaisamy:2016,ChenWangDeng:2018}, and the time-changed stochastic processes are important models in many fields, such as, biology \cite{GoldingCox:2006}, physics \cite{NezhadhaghighiRajabpourRouhani:2011}, ecology \cite{ScherMargolinMetzlerKlafterBerkowitz:2002}, etc.

In fact, the coupled Langevin equation \eqref{free1} describing the subdiffusion dynamics can also be rewritten into a single Langevin equation in physical time $t$ with an additive compound noise $\overline{\xi}(t)=\int_0^{+\infty}\xi(\tau)\delta(t-t(\tau))d\tau$ \cite{CairoliBaule:2015_2}:
\begin{equation}\label{free2}
\dot{y}(t)=\sqrt{2\sigma}\overline{\xi}(t).
\end{equation}
Besides the discussions on the models \eqref{free1} or \eqref{free2}, there are some research works for them with  external forces  \cite{FedotovKorabel:2015,EuleFriedrichJenkoKleinhans:2007,BurovMetzlerBarkai:2010,CairoliBaule:2015_2,CairoliBaule:2015,DieterichKlagesChechkin:2015,MagdziarzWeronKlafter:2008,WeronMagdziarz:2008,ChenWangDeng:2018},
which mainly presented the asymptotic expression of the MSD of the stochastic process for long times, depending on the one-point probability density function (PDF) of the stochastic process. The studies on more general statistical quantities, such as, the correlation coefficient, which reflects the correlation of positions at two different times, as well as  the time averaged MSD, are sometimes ignored. These statistical quantities are significantly important to distinguish the processes with the same diffusion behavior.

In this paper, we investigate the influence of three kinds of common external forces --- position-dependent force, constant force, and time-dependent force. These forces may act on the original process $x(s)$ in \eqref{free1} or on the subordinated process $y(t)$ in \eqref{free2} for different physical realities. These two acting patterns are, respectively, for the cases, where the external force only modifies the dynamical
behavior at the moments of jump or exerts effect for the whole time. The comparisons are made for various effects exerted by different acting patterns with different external forces through some common statistical quantities, such as, ensemble and time averaged MSD, correlation coefficient, and  ergodicity breaking parameter. These quantities mainly depend on the two-point joint PDF of the observed processes, except the ensemble averaged MSD. For different patterns of the force acting on the Langevin equation \eqref{free1} or \eqref{free2}, the methods of obtaining the position autocorrelation function are different, which are fully demonstrated in this paper.

One interesting finding is that the position-independent force acting on the subordinated process in \eqref{free2} does not change the diffusion behavior, ergodic property, and the correlation coefficient, while the position-dependent external force does. But if acting on the original process in \eqref{free1}, the external forces (position-dependent or position-independent) produce different results for almost all the statistical quantities, compared with the ones of free particles. 
In addition, the exponent of ergodicity breaking parameter does not depend on the forces and the acting patterns.

The Fokker-Planck equations govern the PDF $p(y,t)$; generally, they vary with the change of the processes described by the Langevin equations \eqref{free1} and \eqref{free2} with forces.
For the equations, it is found that the Riemann-Liouville fractional derivative with respect to time is included when the external force affects the process only at the moments of jump, while the fractional substantial derivative and another kind of novel fractional derivative are needed when the external force acts on the system for the whole time.

The structure of this paper is as follows. In Sec. \ref{two}, we review the subordinator as well as the inverse subordinator, and briefly present the method of subordination we mainly use.
Then we consider the effects of position-dependent force (harmonic potential), constant force (linear potential), and time-dependent force, respectively, in Sec. \ref{three}--\ref{five}.
The potential properties of the diffusion behaviors are revealed through various statistical quantities. Finally, we make the summaries in Sec. \ref{six} and the detailed derivations of some of the results of the paper are presented in Appendix.

\section{Subordinator}\label{two}
Subordinator is a non-decreasing L\'{e}vy process with stationary and independent increments  \cite{Applebaum:2009} and it can be regarded as a stochastic model of time evolution. The subordinator $t(s)$ in this paper is taken to be $\alpha$-stable one with $0<\alpha<1$ \cite{Applebaum:2009}, which has the characteristic function $\langle\textrm{e}^{-\lambda t(s)}\rangle=\textrm{e}^{-s\lambda^\alpha}$. The brackets $\langle\cdots\rangle$ denote the statistical average over stochastic realizations. The corresponding inverse process, called inverse $\alpha$-stable subordinator $s(t)$ \cite{KumarVellaisamy:2015,AlrawashdehKellyMeerschaertScheffler:2017}, is the first-passage time of the subordinator $\{t(s),\,s\geq 0\}$ , defined as
\begin{equation}
s(t)=\inf_{s>0}\{s:t(s)>t\}.
\end{equation}
In addition, we denote the PDF of the inverse $\alpha$-stable subordinator $s(t)$ as $h(s,t)$ and its Laplace transform ($t\rightarrow\lambda$) $h(s,\lambda)$ is \cite{BauleFriedrich:2005}
\begin{equation}\label{hslambda}
\mathcal{L}_{t\rightarrow \lambda} [h(s,t)]=\int_0^\infty e^{-\lambda t} h(s, t)dt=\lambda^{\alpha-1}e^{-s\lambda^\alpha}.
\end{equation}

The PDF $p(y, t)$ of the subordinated process $y(t):=x(s(t))$ can be written as \cite{BauleFriedrich:2005,Barkai:2001,ChenWangDeng:2018}
\begin{equation}\label{PDF_subor}
p(y, t)=\int_0^\infty  p_0(y,s)h(s,t) ds,
\end{equation}
where $p_0(x,s)$ is {the} PDF of the original process $x(s)$. The moments of the subordinated process $y(t)$ can be obtained through the relation
\begin{equation}\label{relation}
\mathcal{L}_{t\rightarrow \lambda}\langle y^n(t)\rangle=\lambda^{\alpha-1}\mathcal{L}_{s\rightarrow \lambda^\alpha}\langle x^n(s)\rangle
\end{equation}
in Laplace space. Similarly, the two-point joint PDF $p(y_2,t_2;y_1,t_1)$ of $y(t)$ can be obtained through the two-point joint PDF $p_0(x_2,s_2;x_1,s_1)$ of the original stochastic process $x(s)$,
\begin{equation}
\begin{split}
&p(y_2,t_2;y_1,t_1)\\
&~~=\int_0^\infty \int_0^\infty p_0(y_2,s_2;y_1,s_1)h(s_2,t_2;s_1,t_1)ds_1ds_2,
\end{split}
\end{equation}
where $h(s_2,t_2;s_1,t_1)$ is the two-point joint PDF of the inverse subordinator $s(t)$. The correlation function of $y(t)$ in Laplace space ($t_1\rightarrow\lambda_1$, $t_2\rightarrow\lambda_2$) is
\begin{equation}
\begin{split}
&\langle y(\lambda_1)y(\lambda_2)\rangle\\
&~~=\int_0^\infty \int_0^\infty\langle x(s_1)x(s_2)\rangle h(s_2,\lambda_2;s_1,\lambda_1)ds_1 ds_2,
\end{split}
\end{equation}
with \cite{BauleFriedrich:2005}
\begin{equation}
\begin{split}
& h(s_2,\lambda_2;s_1,\lambda_1)\\
&=\delta(s_2-s_1)\frac{\lambda_1^\alpha-(\lambda_1+\lambda_2)^\alpha+\lambda_2^\alpha}{\lambda_1\lambda_2}e^{-s_1(\lambda_1+\lambda_2)^\alpha}\\
&+\Theta(s_2-s_1)\frac{\lambda_2^\alpha[(\lambda_1+\lambda_2)^\alpha-\lambda_2^\alpha]}{\lambda_1\lambda_2}\\
& \cdot e^{-(\lambda_1+\lambda_2)^\alpha s_1}e^{-\lambda_2^\alpha(s_2-s_1)}+\Theta(s_1-s_2)
\\
& \cdot
\frac{\lambda_1^\alpha[(\lambda_1+\lambda_2)^\alpha-\lambda_1^\alpha]}{\lambda_1\lambda_2}e^{-(\lambda_1+\lambda_2)^\alpha s_2}e^{-\lambda_1^\alpha(s_1-s_2)}.
\end{split}
\end{equation}

Based on the formulae above, we have the MSD of the stochastic process $y(t)$ in \eqref{free1} \cite{MetzlerKlafter:2000}
\begin{equation}\label{MSD-free}
  \langle y^2(t)\rangle=\frac{2\sigma}{\Gamma(1+\alpha)}t^\alpha
\end{equation}
and the autocorrelation function \cite{BauleFriedrich:2005} $\langle y(t_1)y(t_2)\rangle=\frac{2\sigma}{\Gamma(1+\alpha)}t_1^\alpha$ for $t_1\leq t_2$. In addition, the time averaged MSD is \cite{HeBurovMetzlerBarkai:2008,LubelskiSokolovKlafter:2008} \begin{equation}\label{TA-free-S}
  \overline{\delta^2(\Delta)}\simeq\frac{2\sigma}{\Gamma(1+\alpha)}\Delta T^{\alpha-1}
\end{equation}
for $\Delta\ll T$. The corresponding Fokker-Planck equation, governing the PDF $p(y,t)$ of finding the particle at position $y$ at time $t$, is \cite{MetzlerKlafter:2000,MetzlerKlafter:2004}
\begin{eqnarray}\label{FP_free}
\frac{\partial p(y,t)}{\partial t}=\sigma  \frac{\partial^2}{\partial y^2}D_t^{1-\alpha}p(y,t).
\end{eqnarray}
The symbol $D_t^{1-\alpha}$ is the Riemann-Liouville fractional derivative \cite{Podlubny:1999}, defined as
\begin{eqnarray}
D_t^{1-\alpha}p(y,t)=\frac{1}{\Gamma(\alpha)}\frac{\partial}{\partial t}\int_0^t (t-t')^{\alpha-1}p(y,t')dt',
\end{eqnarray}
which is a nonlocal time derivative and indicates the non-Markovian property of the process $y(t)$.

\section{Subdiffusive dynamics in harmonic potential}\label{three}
In the following two subsections, we respectively discuss two cases: acting on the Langevin equations \eqref{free1} and \eqref{free2} by the harmonic potential.
By comparing some statistical quantities, including ensemble and time averaged MSD, correlation coefficient, and ergodicity breaking parameter, we find some significant differences and interesting phenomena, especially in the latter case where the position-dependent external force acts on the system all the time.

\subsection{Force acting on original process $x(s)$}
Consider the Langevin system with a harmonic potential on the original process $x(s)$ \cite{EuleFriedrichJenkoKleinhans:2007,BurovMetzlerBarkai:2010}
\begin{equation}\label{x_harmonic}
\dot{x}(s)=-\gamma x(s)+\sqrt{2\sigma}\xi(s), \qquad \dot{t}(s)=\eta(s),
\end{equation}
where $\gamma$ is a positive constant, $\xi(s)$ and $\eta(s)$ are two independent noises defined in \eqref{free1}. The harmonic potential $V(x)=\gamma x^2/2$ leads to a friction-like force $F(x)=-dV(x)/dx=-\gamma x$ in the first equation of \eqref{x_harmonic}. Based on \eqref{x_harmonic}, a new single Langevin equation in physical time $t$ of the subordinated process $y(t)=x(s(t))$ can be obtained as
\begin{equation}\label{y_harmonic}
\dot{y}(t)=-\gamma y(t)\dot{s}(t)+\sqrt{2\sigma}\overline{\xi}(t),
\end{equation}
with $\overline{\xi}(t)=\int_0^{+\infty}\xi(\tau)\delta[t-t(\tau)]d\tau$, or equivalently, $\overline{\xi}(t)=\xi(s(t))\dot{s}(t)$, since
\begin{equation}
\begin{split}
y(t)&=x(s(t))\\
&=-\gamma \int_0^{s(t)} x(s')ds'+\sqrt{2\sigma}\int_0^{s(t)} \xi(s')ds'\\
&=-\gamma \int_0^t x(s(\tau))ds(\tau)+\sqrt{2\sigma}\int_0^t \xi(s(\tau))ds(\tau)\\
&=-\gamma \int_0^t y(\tau)ds(\tau)+\sqrt{2\sigma}\int_0^t \xi(s(\tau))ds(\tau).
\end{split}
\end{equation}
The noise $\overline{\xi}(t)$ here can be regarded as {the formal derivative of the} time-changed Brownian motion $B(s(t))$.
The external force in \eqref{x_harmonic} only changes the motion of the particles at the instant of jumps; in fact, this mechanism can be easily found from the equivalent Langevin equation in physical time \eqref{y_harmonic}, i.e., when a particle suffers a trapping event before next jump, the internal time process $s(t)$ remains a constant and the external force becomes zero due to $\dot{s}(t)=0$ in \eqref{y_harmonic}.

\begin{figure}
\begin{minipage}{0.35\linewidth}
  \centerline{\includegraphics[scale=0.38]{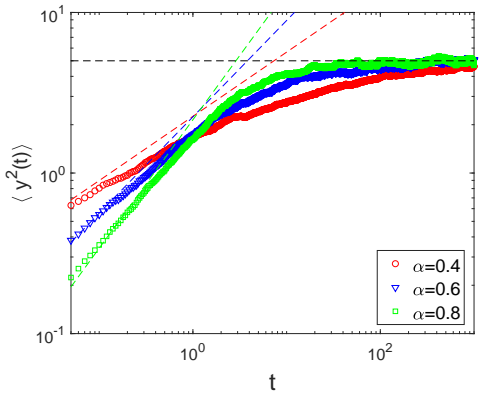}}
  \centerline{}
  \end{minipage}
\caption{Simulation results of the MSD of stochastic process \eqref{x_harmonic} for different $\alpha$. Color markers represent the simulation results of MSD with parameters $\sigma=1$ and $\gamma=0.2$ averaging over $2000$ trajectories. Color dashed lines and black dashed lines represent the asymptotic theoretical values of MSD for short and long times, seeing \eqref{MSD-1} and \eqref{longtime}, respectively.
}\label{MSD1_1}
\end{figure}

Using formula \eqref{relation}, it can be got that the first moment of the stochastic process $y(t)$ is zero due to symmetry and the MSD is
\begin{equation}
\langle y^2(t)\rangle=\frac{\sigma}{\gamma}-\frac{\sigma}{\gamma}E_\alpha(-2\gamma t^\alpha),
\end{equation}
by utilizing $\langle x^2(s)\rangle=\frac{\sigma}{\gamma}(1-e^{-2\gamma s})$ with the initial position $x_0=0$. Considering the asymptotic expression of the Mittag-Leffler function \cite{Erdelyi:1981} for small $t$: $E_\alpha(-2\gamma t^\alpha)\simeq1-\frac{2\gamma t^\alpha}{\Gamma(1+\alpha)}$, the asymptotic form of the MSD for short times $t\ll(2\gamma)^{-\frac{1}{\alpha}}$ is
\begin{equation}\label{MSD-1}
\langle y^2(t)\rangle\simeq\frac{2\sigma}{\Gamma(1+\alpha)}t^\alpha,
\end{equation}
which coincides with the MSD of a free particle in \eqref{MSD-free} and implies that the harmonic potential does not affect the diffusion dynamics in short times. But for long times $t\gg(2\gamma)^{-\frac{1}{\alpha}}$, one has
\begin{equation}\label{longtime}
\langle y^2(t)\rangle\simeq \frac{\sigma}{\gamma}-\frac{\sigma}{2\gamma^2\Gamma(1-\alpha)}t^{-\alpha}
\end{equation}
because of the asymptotic expression $E_\alpha(-2\gamma t^\alpha)\simeq \frac{t^{-\alpha}}{2\gamma \Gamma(1-\alpha)}$ for large $t$. The saturation plateau value, denoted as $\langle y^2\rangle_{\textrm{th}}=\frac{\sigma}{\gamma}$, is approached at the power-law rate. The simulation results for different $\alpha$ are shown in Fig. \ref{MSD1_1}. It can be seen that the MSD with a smaller $\alpha$ tends to the saturation plateau value more slowly, being an expected dynamical behavior within a confined harmonic potential due to smaller $\alpha$ corresponding to longer waiting time. This process behaves as a localization diffusion for long times. Compared with the original process $x(s)$, the MSD of which relaxes to the value $\frac{\sigma}{\gamma}$ exponentially, the subordinator $s(t)$ in this model only changes the convergence rate but keeps the same saturation plateau value.

In addition,  the position autocorrelation function is \cite{BurovMetzlerBarkai:2010}
\begin{equation}
\langle y(t_1)y(t_2)\rangle\simeq \frac{\sigma}{\gamma}\frac{B(t_1/t_2, \alpha, 1-\alpha)}{\Gamma(\alpha)\Gamma(1-\alpha)}
\end{equation}
for large $t_1$, $t_2$ ($t_2\geqslant t_1$), where $B(z, a, b)$ is the incomplete Beta function \cite{AbramowitzStegun:1972}. Therefore, this process is non-stationary since the position autocorrelation function cannot be expressed as a function of time difference $|t_1-t_2|$.
For fixed $t_1$ and $t_2\rightarrow \infty$, the correlation coefficient $r[y(t_1),y(t_2)]$ of the stochastic process $y(t)$, which characterizes the correlation of position at two different times, can be obtained as
\begin{equation}\label{co1}
\begin{split}
r[y(t_1),y(t_2)]&=\frac{\langle (y(t_1)-\langle y(t_1)\rangle)(y(t_2)-\langle y(t_2)\rangle) \rangle}{\sqrt{\langle (y(t_1)-\langle y(t_1)\rangle)^2\rangle\langle(y(t_2)-\langle y(t_2)\rangle)^2\rangle }}\\
&\simeq \frac{1}{\Gamma(1-\alpha)\Gamma(1+\alpha)}\left(\frac{t_1}{t_2}\right)^\alpha,
\end{split}
\end{equation}
where we have used the asymptotic expression of the incomplete Beta function for small $z$, i.e., $B(z,a,b)\simeq z^a/a$.
It can be seen that the process $y(t)$ described by model \eqref{x_harmonic} is long-range dependent due to the power-law decay with $0<\alpha<1$ of the correlation coefficient in \eqref{co1}.
On the other hand, as a result of the harmonic potential, the correlation becomes weaker than that of a free particle, the correlation coefficient of which  is
\begin{equation}\label{CC-free}
r_0[y(t_1),y(t_2)]=\left(\frac{t_1}{t_2}\right)^\frac{\alpha}{2}.
\end{equation}

The Fokker-Planck equation corresponding to the Langevin equation \eqref{x_harmonic} is \cite{MetzlerBarkaiKlafter:1999,MetzlerKlafter:2000,GajdaMagdziarz:2010},
\begin{eqnarray}\label{FP}
\frac{\partial p(y,t)}{\partial t}=\mathcal{L}_\textrm{FP}D_t^{1-\alpha}p(y,t)
\end{eqnarray}
with the Fokker-Planck operator $\mathcal{L}_\textrm{{FP}}=-\frac{\partial}{\partial y} F(y)+\sigma\frac{\partial^2}{\partial y^2} $ (here $F(y)=-\gamma y$), which can be derived by three different methods. The first one is based  on the relation \eqref{PDF_subor} between the PDF of subordinated process and original process \cite{ChenWangDeng:2018}. The second one is to take the parameter $p=0$ in the Feynman-Kac equation  \cite{CairoliBaule:2017,WangChenDeng:2018}. As for the last method, (\ref{FP}) can be got from the master equation in CTRW model within a harmonic potential \cite{MetzlerKlafter:2000,CarmiBarkai:2011}.

Let us turn to the time averaged MSD, defined as \cite{MetzlerJeonCherstvyBarkai:2014,BurovJeonMetzlerBarkai:2011,AkimotoCherstvyMetzler:2018,HouCherstvyMetzlerAkimoto:2018}
\begin{equation}\label{TAdefination2}
\begin{split}
  \overline{\delta^2(\Delta)}
  &=\frac{1}{T-\Delta}\int_0^{T-\Delta}  [(y(t+\Delta)-y(t))  \\
  &~~~    -\langle y(t+\Delta)-y(t)\rangle]^2dt,
\end{split}
\end{equation}
where $\Delta$ is the lag time, and $T$ is the measurement time. We emphasize that the lag time $\Delta$ separating the displacement between trajectory points is much shorter than the measurement time $T$.
Sometimes, the time averaged MSD of some stochastic systems confined in a harmonic potential approaches twice the ensemble averaged MSD $\langle y^2\rangle_{\textrm{th}}$ for long times, such as, overdamped Brownian motion, fractional Brownian motion, and fractional Langevin dynamics  \cite{JeonMetzler:2012,JeonLeijnseOddershedeMetzler:2013}.
Different from it, the time averaged MSD of the confined model \eqref{x_harmonic} is sublinear in lag time $\Delta$ \cite{BurovMetzlerBarkai:2010,BurovJeonMetzlerBarkai:2011,NeusiusSokolovSmith:2009}
\begin{equation}\label{TAMSD}
\langle\overline{\delta^2(\Delta)}\rangle\simeq \frac{2\sigma}{\gamma}\frac{\textrm{sin}(\alpha\pi)}{\alpha(1-\alpha)\pi}\left(\frac{\Delta}{T}\right)^{1-\alpha}
\end{equation}
with $T \gg \Delta\gg(1/\gamma_1)^{1/\alpha}$. Here $\gamma_1$ is the smallest nonzero eigenvalue of the Fokker-Planck operator $\mathcal{L}_\textrm{{FP}}$. The disagreement between the ensemble and time averaged MSD, the former is constant $\Delta^0$ and the latter scales as $\Delta^{1-\alpha}$, indicates non-ergodicity of the stochastic process $y(t)$. At short lag times $\Delta$, the linear scaling in lag times is observed \cite{BurovMetzlerBarkai:2010}
$\langle\overline{\delta^2(\Delta)}\rangle\simeq\frac{2\sigma}{\Gamma(1+\alpha)}\Delta T^{\alpha-1}$,
which is the same as the one of a free particle in \eqref{TA-free-S} since the potential has not begun to affect the stochastic process.

Now, we consider the influence of the harmonic potential in terms of the ergodicity breaking parameter, defined as $\mathcal{EB}=\langle\overline{\delta^2}(\Delta)\rangle/\langle x^2(\Delta)\rangle.$
For the process of free particle,
the ergodicity breaking parameter is
\begin{equation}\label{EBfree}
\mathcal{EB}\simeq \left( \frac{T}{\Delta}\right)^{\alpha-1},
\end{equation}
while for the case with harmonic potential, the ergodicity breaking parameter becomes
\begin{equation}
\mathcal{EB}\simeq \frac{2\sin(\alpha\pi)}{\alpha(1-\alpha)\pi}\left( \frac{T}{\Delta}\right)^{\alpha-1}.
\end{equation}
They have the same exponents of $T/\Delta$, but the coefficient of the latter is larger.


\subsection{Force acting on subordinated process $y(t)$}

The external force in \eqref{x_harmonic} only makes an influence on the dynamical behavior at the moments of jump; contrary to it, the force may keep acting on the system all the time, even when the particle is trapped. Recently, such a model has been proposed in \cite{CairoliBaule:2015_2}, where the harmonic potential is assumed to directly act on the subordinated process $y(t)$ in physical times,
\begin{equation}\label{y_harmonic2}
\dot{y}(t)=-\gamma y(t)+\sqrt{2\sigma}\overline{\xi}(t),
\end{equation}
where $\overline{\xi}(t)$ is the same noise as the one in \eqref{y_harmonic}. The two point correlation function of this compound noise $\overline{\xi}(t)$ \cite{CairoliBaule:2015_2} could be gotten through the inverse Laplace transform
\begin{equation}
\begin{split}
\langle \overline{\xi}(t_1)\overline{\xi}(t_2)\rangle&=\mathcal{L}_{\lambda_1\rightarrow t_1, \lambda_2\rightarrow t_2}^{-1}[(\lambda_1+\lambda_2)^{-\alpha}]\\
&=t_1^{\alpha-1}\delta(t_1-t_2)/\Gamma(\alpha).
\end{split}
\end{equation}
The harmonic potential acts as a friction-like force $-\gamma y(t)$; even in the constant period of inverse subordinator $s(t)$,  it still influences the stochastic dynamics.
{Actually, the Langevin equation \eqref{y_harmonic2} can be rewritten as a coupled Langevin system with a subordinator as
\begin{equation}\label{x_harmonic2}
\dot{x}(s)=-\gamma x(s)\eta(s)+\sqrt{2\sigma}\xi(s), \qquad \dot{t}(s)=\eta(s).
\end{equation}
More precisely, the solution of \eqref{y_harmonic2} is
\begin{equation}\label{solution}
  y(t)=\sqrt{2\sigma}\int_0^{t}e^{-\gamma (t-\tau)}\overline{\xi}(\tau)d\tau
\end{equation}
with initial condition $y_0=0$, which is equivalent to
\begin{equation}
x(s)=\sqrt{2\sigma}\int_0^s e^{-\gamma(t(s)-t(\tau))}d B(\tau),
\end{equation}
by replacing $s$ with $s(t)$.
Compared with \eqref{x_harmonic}, the friction term $-\gamma x(s)$ is multiplied by the L\'{e}vy noise $\eta(s)$, which acts as a multiplicative noise in the first equation in \eqref{x_harmonic2}.}

\begin{figure}
\begin{minipage}{0.35\linewidth}
  \centerline{\includegraphics[scale=0.249]{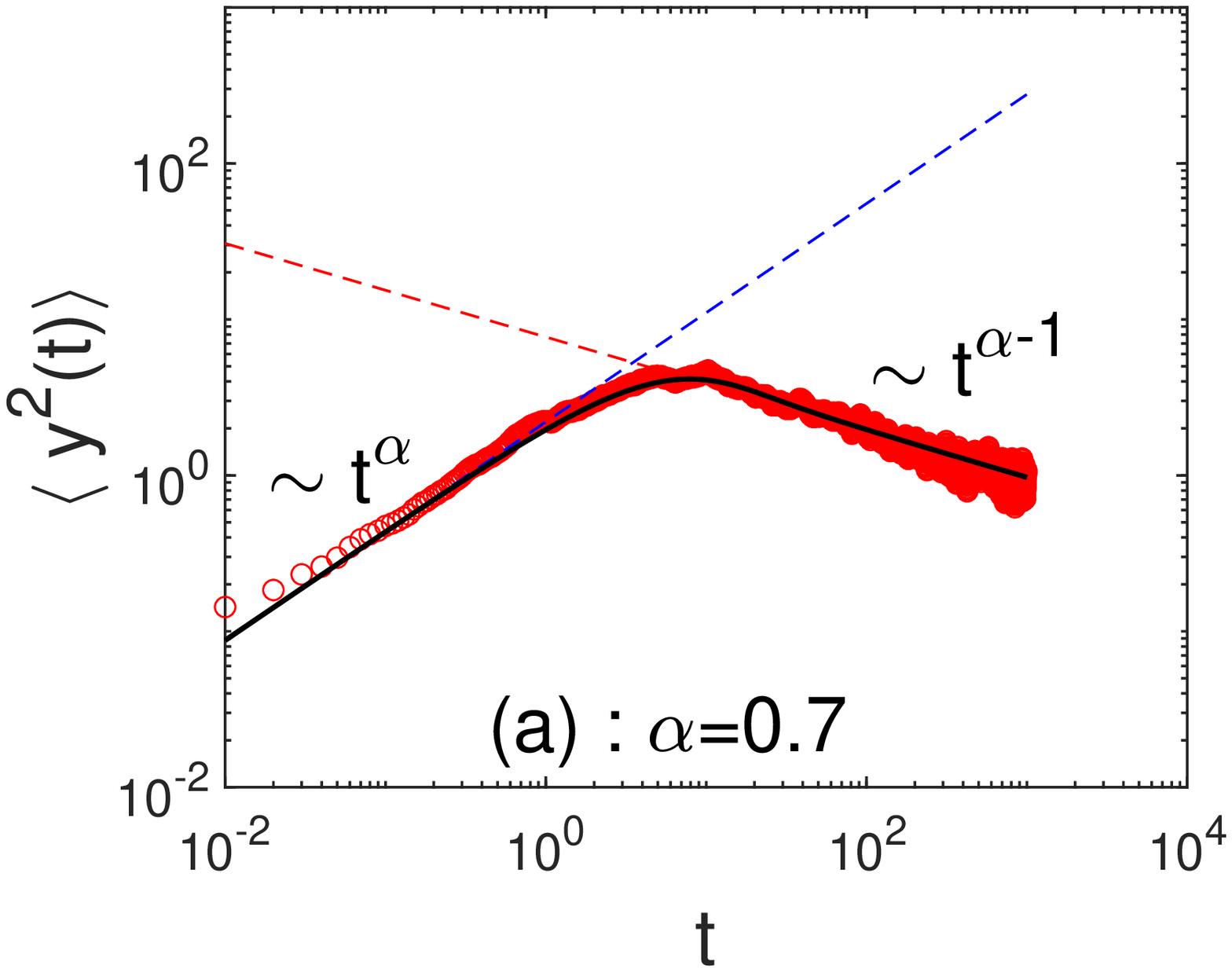}}
  \centerline{}
\end{minipage}
\hspace{1cm}
\begin{minipage}{0.35\linewidth}
  \centerline{\includegraphics[scale=0.249]{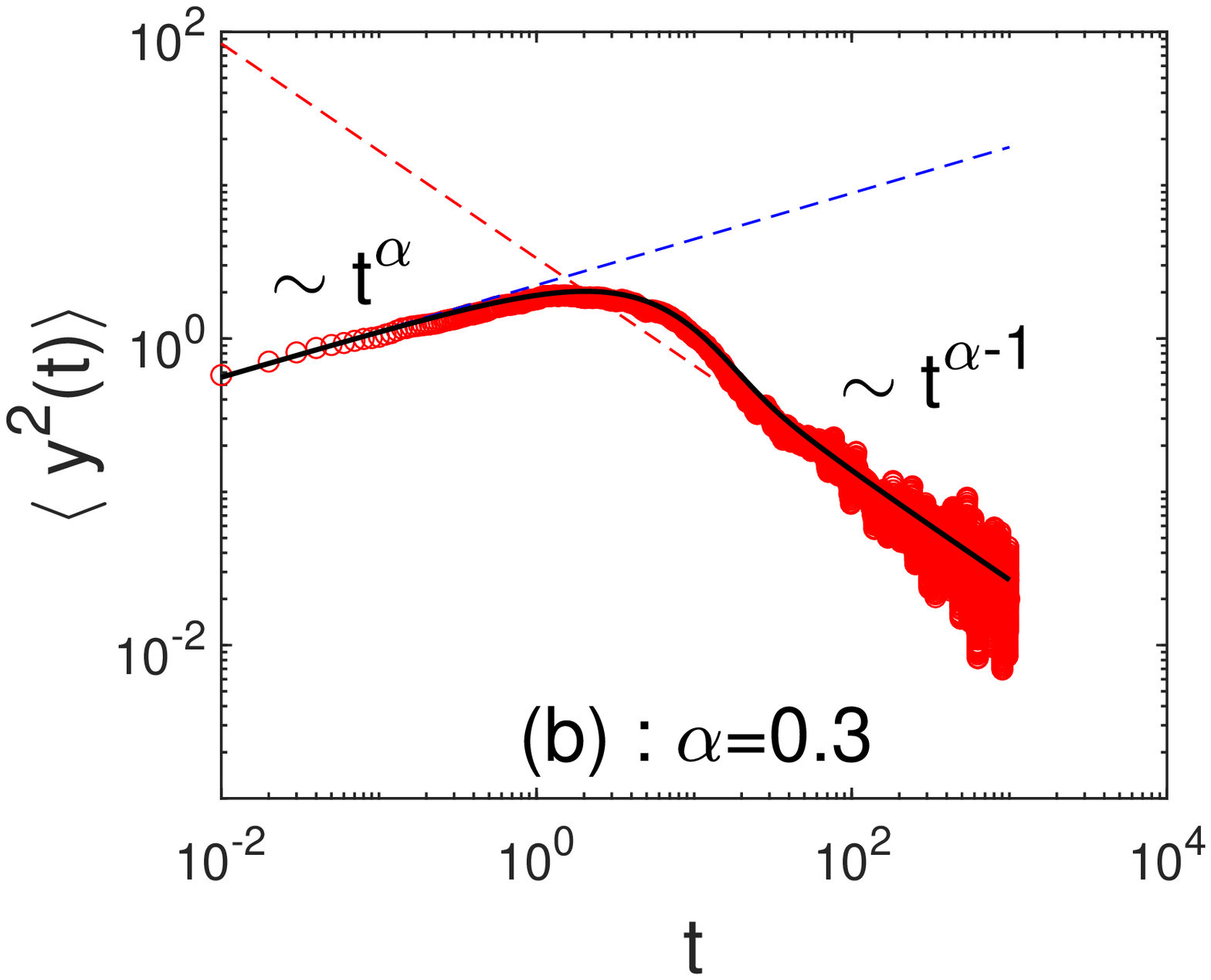}}
  \centerline{}
\end{minipage}
\caption{Simulation results of the MSD of stochastic process described by the Langevin equation \eqref{y_harmonic2} for different $\alpha$. The parameters are, respectively, taken as $\sigma=1$, $\gamma=0.1$, $\alpha=0.7$ (a) or $\alpha=0.3$ (b), and the initial position $y_0=0$. 
Blue dashed lines and the red dashed lines represent the asymptotic theoretical values in \eqref{short2} and \eqref{long2}, respectively, and the black solid lines signify the theoretical values \eqref{MSD_2}. Red circle-markers are the simulation results.  }\label{MSD1_2}
\end{figure}

From \eqref{solution}, it can be calculated that the mean of $y(t)$ is zero and the MSD is
\begin{equation}\label{MSD_2}
\langle y^2(t)\rangle=\frac{2\sigma}{\Gamma(1+\alpha)}e^{-2\gamma t}t^\alpha~_1F_1(\alpha, 1+\alpha; 2\gamma t)
\end{equation}
with the confluent hypergeometric function \cite{AbramowitzStegun:1972} $_1F_1(a, b; z)=\frac{\Gamma(b)}{\Gamma(a)\Gamma(b-a)}\int_0^1e^{zu}u^{a-1}(1-u)^{b-a-1}du$. The asymptotic expansion of MSD for short times $t\ll(2\gamma)^{-1}$ is
\begin{equation}\label{short2}
\langle y^2(t)\rangle\simeq\frac{2\sigma}{\Gamma(1+\alpha)}t^\alpha,
\end{equation}
which coincides to that of the free particle \eqref{MSD-free}. For long times $t\gg(2\gamma)^{-1}$, using the asymptotic expansion $_1F_1(a, b; z)\simeq \Gamma(b)\left( e^z z^{a-b}/\Gamma(a)+(-z)^{-a}/\Gamma(b-a)\right)$ for large $z$ \cite{AbramowitzStegun:1972}, we get
\begin{equation}\label{long2}
\langle y^2(t)\rangle\simeq\frac{\sigma}{\gamma\Gamma(\alpha)}t^{\alpha-1},
\end{equation}
which tends to zero at the power-law rate. The consistency between simulation and the theoretical results about the MSD of model \eqref{y_harmonic2} can be found in Fig. \ref{MSD1_2}.

Different from the model \eqref{x_harmonic}, the subordinator in this model changes not only the convergence rate but also the stationary value of MSD for long times. The external force in this model damps the oscillation of the particle in harmonic potential and drags it towards zero for all times; while the subordinated process \eqref{x_harmonic} does not get dragged to zero position during waiting times since the external force is zero during these time periods.

Let us pay attention to the critical time distinguishing two different scales in two models \eqref{x_harmonic} and \eqref{y_harmonic2}. It is $t=(2\gamma)^{-\frac{1}{\alpha}}$ in the first model, depending on the parameter $\alpha$ and influenced by the inverse subodinator $s(t)$. On the contrary, the critical time is $t=(2\gamma)^{-1}$ in the second model, which is independent of $\alpha$ and as same as that of original process $x(s)$ in \eqref{x_harmonic}. It means that the critical time  in the second model is independent of the subordinator $s(t)$ and fully determined by the harmonic potential itself. On the other hand, the {size relation} between these two critical time is uncertain, depending on $\gamma$. If $\gamma<\frac{1}{2}$, the time during which harmonic potential does not work is longer in \eqref{x_harmonic} than in model \eqref{y_harmonic2}.

Taking Laplace transform in \eqref{y_harmonic2} yields $\langle y(\lambda_1)y(\lambda_2)\rangle=\frac{2\sigma}{(\lambda_1+\gamma)(\lambda_2+\gamma)(\lambda_1+\lambda_2)^\alpha}$, from which, one arrives at the position autocorrelation function
\begin{equation}\label{correlation_2}
\langle y(t_1)y(t_2)\rangle=\frac{2\sigma}{\Gamma(1+\alpha)}e^{-\gamma (t_1+t_2)}t_1^\alpha~_1F_1(\alpha, 1+\alpha; 2\gamma t_1)
\end{equation}
for $t_2\geq t_1$. This position autocorrelation function shows the process described by model \eqref{y_harmonic2} is non-stationary, as well as the process in model \eqref{y_harmonic}. Then using the MSD \eqref{MSD_2} and  autocorrelation function \eqref{correlation_2} of process $y(t)$, one obtains the correlation coefficient of the stochastic process $y(t)$ for fixed $t_1$ and $t_2\rightarrow \infty$
\begin{equation}
\begin{split}
r[y(t_1),y(t_2)]\simeq [G_1(t_1)]^{\frac{1}{2}}e^{-\gamma t_2}t_2^{\frac{1-\alpha}{2}}
\end{split}
\end{equation}
with $G_1(t_1)=2\gamma t_1^\alpha/\alpha\cdot {}_1F_1(\alpha, 1+\alpha; 2\gamma t_1)$ being a constant for fixed $t_1$. Different from \eqref{co1}, the correlation coefficient here exponentially decays to zero. It means that the positions at two different times are no longer long-range dependent as a result of the continuous influence of harmonic potential in model \eqref{y_harmonic2}.

The ensemble averaged integrand in \eqref{TAdefination2} could be obtained as
\begin{equation}
\begin{split}
&\langle [y(t+\Delta)-y(t)]^2\rangle  \\
&~=\frac{2\sigma}{\Gamma(1+\alpha)}\Big(e^{-2\gamma (t+\Delta)}(t+\Delta)^\alpha{}
_1F_1(\alpha, 1+\alpha; 2\gamma (t+\Delta))\\
&~~~~+(1-2e^{-\gamma\Delta})e^{-2\gamma t}t^\alpha~_1F_1(\alpha, 1+\alpha; 2\gamma t)\Big).
\end{split}
\end{equation}
This result not only depends on the lag time $\Delta$ but also the time $t$, implying the aging phenomenon by regarding $t$ as the aging time $t_a$ in a system. It means that the observation time impacts the statistical quantities of a system, which was initially prepared. After some calculations, the ensemble-time averaged MSD is obtained,
\begin{equation}
\begin{split}
&\langle \overline{\delta^2(\Delta)}\rangle=\frac{2\sigma}{\Gamma(1+\alpha)(T-\Delta)} \\
&~~~~~\cdot \Big[M(T)-M(\Delta)+(1-2e^{-\gamma \Delta})M(T-\Delta)\Big].
\end{split}
\end{equation}
Here, $M(\Delta)=\frac{\Delta^{\alpha+1}}{\alpha+1}~_2F_2(\alpha+1, 1; \alpha+2, \alpha+1; -2\gamma \Delta)$ and $_2F_2(a,b;c,d; z)$ is the hypergeometric function \cite{AbramowitzStegun:1972}.

\begin{figure}
\begin{minipage}{0.35\linewidth}
  \centerline{\includegraphics[scale=0.4]{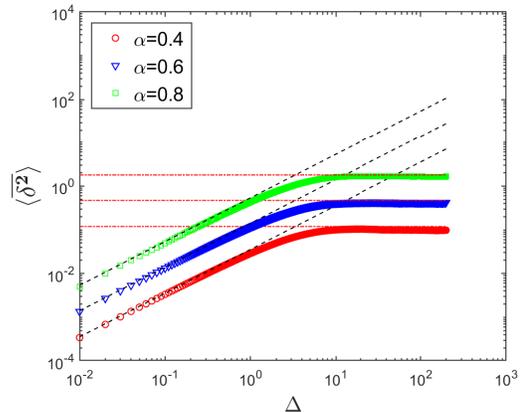}}
  \end{minipage}
\caption{Simulation results of the ensemble-time averaged MSD of stochastic process described by the Langevin equation \eqref{y_harmonic2} for different $\alpha$. The parameters are taken as $\sigma=1$, $\gamma=0.3$, and the initial position $y_0=0$. The  measurement time $T=1000$. Black dashed lines and red dashed dot lines represent the asymptotic theoretical results for short lag time and large lag time in \eqref{TA-free-S} and \eqref{TA},  respectively. Color markers are the simulation results about the ensemble-time averaged MSD over $500$ trajectories. }\label{ergo1_2}
\end{figure}

For short $\Delta$, i.e., $\Delta \ll\gamma^{-1}$, the time averaged MSD is the same as the one for a free particle in \eqref{TA-free-S}, growing linearly with the lag time.
For large $\Delta$, i.e., $\gamma^{-1}\ll\Delta\ll T$, by using the asymptotic expression of H-function \cite{MathaiSaxenaHaubold:2009}, we have
\begin{equation}
\begin{split}
_2F_2&(\alpha+1, 1; \alpha+2, \alpha+1; -2\gamma \Delta)\\
&=\Gamma(\alpha+2)H_{2,3}^{1,2}\left[2\gamma \Delta  \left|
\begin{array}{l}
  (-\alpha,1),(0,1)  \\   (0,1),(-1-\alpha,1),(-\alpha,1)
\end{array}
\right.\right]  \\
&\simeq(\alpha+1)(2\gamma)^{-1}\Delta^{-1},
\end{split}
\end{equation}
and find that the time averaged MSD approaches to a constant
\begin{equation}\label{TA}
\langle \overline{\delta^2(\Delta)}\rangle\simeq\frac{2\sigma}{\gamma\Gamma(1+\alpha)} T^{\alpha-1},
\end{equation}
which is different from the time averaged MSD \eqref{TAMSD} in model \eqref{y_harmonic}. The disagreement between the ensemble and time averaged MSD, scaling as $\Delta^{\alpha-1}$ and $\Delta^0$ respectively,  means the non-ergodicity of this system. Fig. \ref{ergo1_2} shows the consistency of the simulation results and analytical ones of the ensemble-time averaged MSD for different $\alpha$. It also can be found that the turning point is almost $\gamma^{-1}$ in Fig. \ref{ergo1_2}, beyond which the plateau value \eqref{TA} is approached. The ergodicity breaking parameter of this model is
\begin{equation}
\mathcal{EB}\simeq\frac{2}{\alpha}\left(\frac{T}{\Delta}\right)^{\alpha-1},
\end{equation}
which is also similar to the one of free particle in \eqref{EBfree} but with a larger coefficient.

\section{Subdiffusive dynamics in linear potential}\label{four}
This section focuses on the influence of the linear potential acting on the original process in \eqref{free1} and on the subordinated process in \eqref{free2}.

\subsection{Force acting on original process $x(s)$}
We first consider the Langevin system with linear potential, i.e., a constant external force \cite{CairoliBaule:2015,DieterichKlagesChechkin:2015,ChenWangDeng:2018}
\begin{equation}\label{x_constant}
\dot{x}(s)=F+\sqrt{2\sigma}\xi(s), \qquad \dot{t}(s)=\eta(s).
\end{equation}
The corresponding single Langevin equation of $y(t)=x(s(t))$ in physical time is
\begin{equation}\label{y_constant}
\dot{y}(t)=F\dot{s}(t)+\sqrt{2\sigma}\overline{\xi}(t),
\end{equation}
which evidently shows that the external force only acts at the moments of jump and does not affect the particle during waiting times (trapping events). In addition, the Fokker-Planck equation with respect to this Langevin system is also \eqref{FP} by replacing $F(y)$ with $F$.

\begin{figure}
\begin{minipage}{0.35\linewidth}
  \centerline{\includegraphics[scale=0.248]{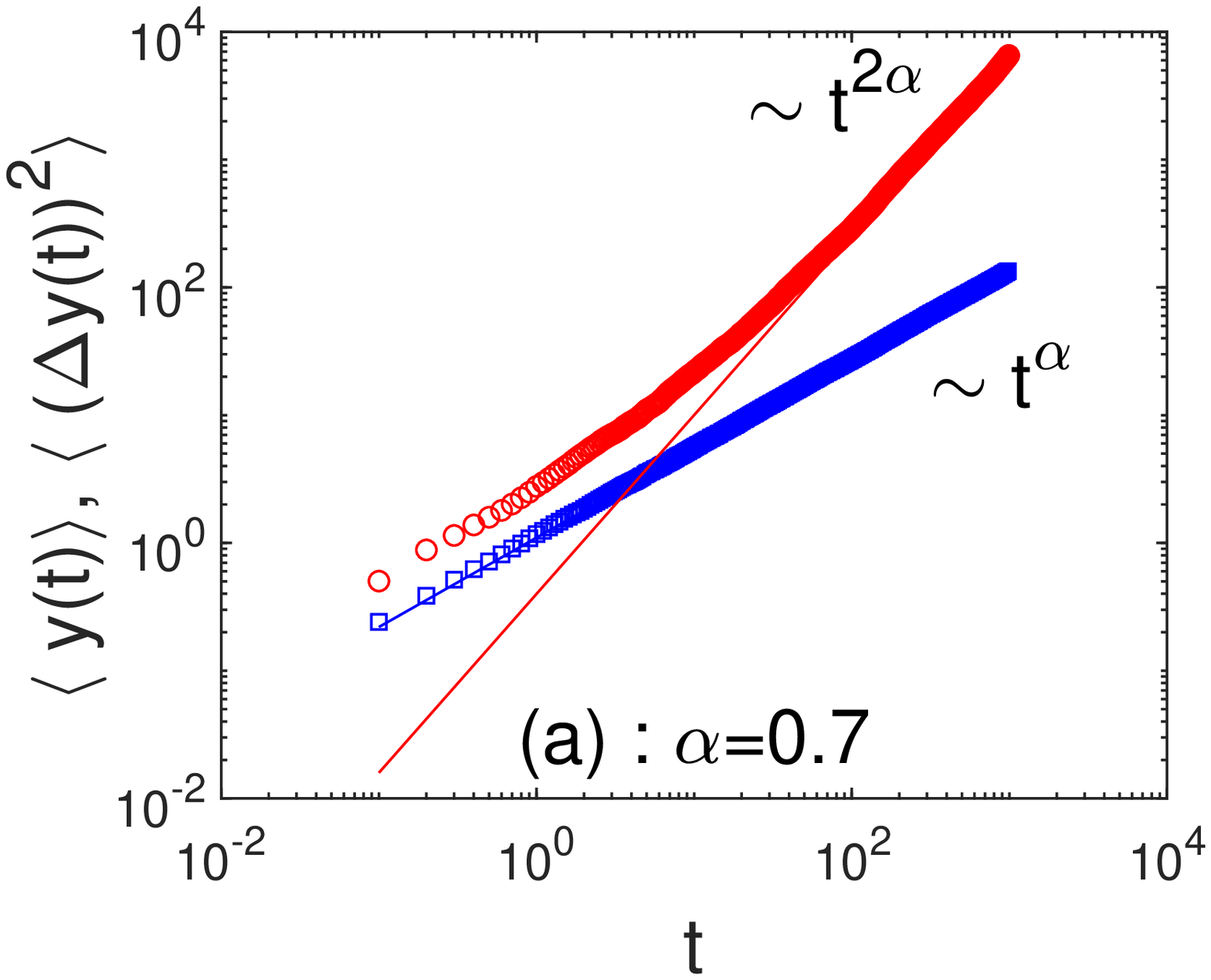}}
  \centerline{}
\end{minipage}
\hspace{1cm}
\begin{minipage}{0.35\linewidth}
  \centerline{\includegraphics[scale=0.248]{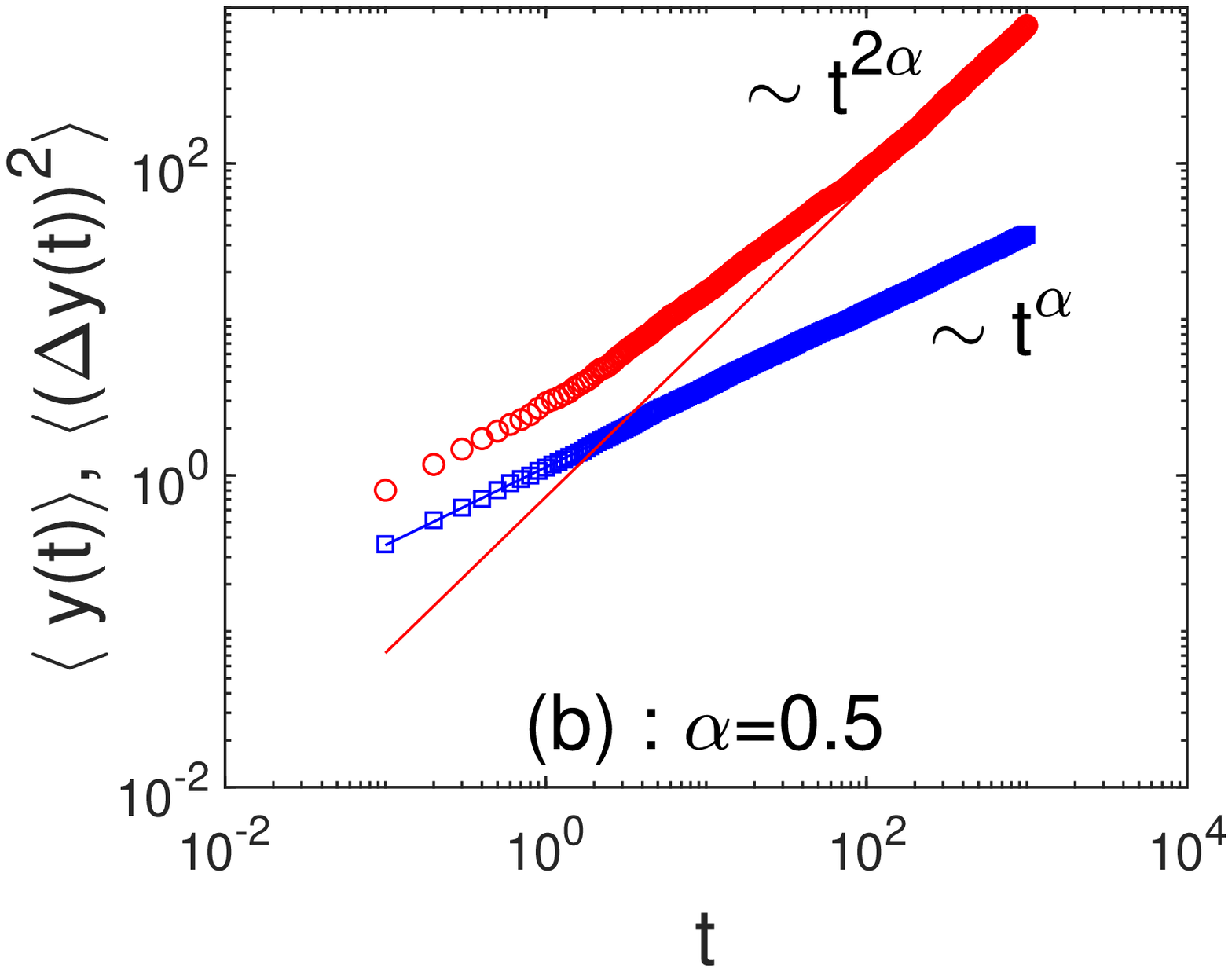}}
  \centerline{}
\end{minipage}
\caption{Simulation results of the MSD of stochastic process described by the Langevin equation \eqref{x_constant} for different $\alpha$. The parameters are, respectively, taken as $\sigma=1$, $F=1$, $\alpha=0.7$ (a) or $\alpha=0.5$ (b), and the initial position $y_0=0$. 
Blue solid lines and the blue square-markers represent
the theoretical value in \eqref{liangge} and the simulation result of the first moment. Besides, the red solid lines and the red circle-markers are, respectively, the theoretical value in \eqref{liangge} and the simulation result
of the ensemble averaged MSD.  }\label{MSD2_1}
\end{figure}

With the procedures similar to the case of harmonic potential in Section \ref{three}, the first moment and MSD of the subordinated process $y(t)=x(s(t))$ are obtained as \cite{CompteMetzlerCamacho:1997,MetzlerKlafter:2000,DieterichKlagesChechkin:2015,ChenWangDeng:2018}
\begin{equation}\label{liangge}
\begin{split}
\langle y(t) \rangle&=\frac{F}{\Gamma(1+\alpha)}t^\alpha,  \\
\langle(\Delta y(t))^2\rangle&=\left( \frac{F^2}{\alpha\Gamma(2\alpha)}-\frac{F^2}{\alpha^2\Gamma^2(\alpha)}\right)t^{2\alpha}
\end{split}
\end{equation}
with $0<\alpha<1$. The simulation results are presented in Fig. \ref{MSD2_1}. This subordinated Langevin system shows subdiffusion when $0<\alpha<\frac{1}{2}$ and superdiffusion when $\frac{1}{2}<\alpha<1$.  It is more or less interesting that the waiting time with infinite mean value produces superdiffusion. Compte \emph{et al.} \cite{CompteMetzlerCamacho:1997} explained this phenomenon that some stagnated particles are not continuously dragged by the stream and thus slow down the advancement of the center of mass of the particles, hence the main dispersion mechanism should be convection.
The deviation of the MSD of Langevin systems with constant force from the one of free particle implies this external force is a biasing force.
In addition, the generalised Einstein relation \cite{BouchaudGeorges:1990,MetzlerBarkaiKlafter:1999,BarkaiMetzlerKlafter:2000,MetzlerKlafter:2000,BlickleSpeckLutzSeifertBechinger:2007} connects the first moment of the particle displacements under a constant force to the second moment of a free particle,  $\langle y(t)\rangle_F=\frac{F}{2k_B\mathcal{T}}\langle y^2(t)\rangle_0$. Here $k_B$ is the Boltzman constant and $\mathcal{T}$ is absolute temperature. Here we emphasize that the generalised Einstein relation holds for a subordinated process if it is valid for the original process. More precisely, the subordinator here affects the moments, simultaneously, as
\begin{equation}
\begin{split}
\langle y(t)\rangle_F&=\int_0^\infty\langle x(s)\rangle_Fh(s,t)ds\\
&=\frac{F}{2k_B\mathcal{T}}\int_0^\infty\langle x^2(s)\rangle_0h(s,t)ds\\
&=\frac{F}{2k_B\mathcal{T}}\langle y^2(t)\rangle_0.
\end{split}
\end{equation}

\begin{figure}
\begin{minipage}{0.35\linewidth}
  \centerline{\includegraphics[scale=0.38]{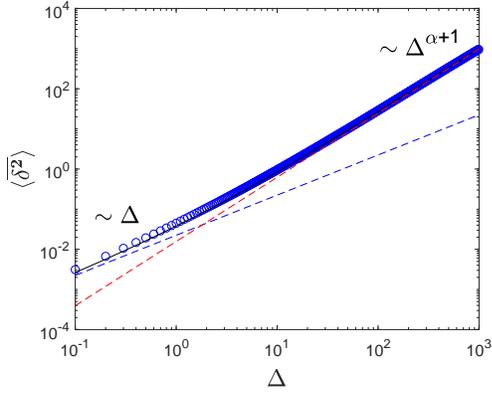}}
  \end{minipage}
\caption{Ensemble-time averaged MSD of the stochastic process described by Langevin equation \eqref{x_constant}. Black solid line shows the theoretical result \eqref{TA_F}, which coincides with the simulation result of the ensemble-time averaged MSD, represented by the blue circle-markers. In addition, the red dashed line and the blue dashed line are the asymptotic expressions of the time averaged MSD for long and short lag times, scaled as $\Delta^{\alpha+1}$ and $\Delta$, respectively. Parameter values: $T=10^5$, $F=1$, $\sigma=1$, and $\alpha=0.6$. Here we choose a big $\alpha$ to see the transition of the time averaged MSD more clearly.}\label{ergo2_1}
\end{figure}

Let us now see the ergodicity of model \eqref{x_constant}. By using the technique of Laplace transform, we get the position autocorrelation function of $y(t)$ for $t_2\geq t_1$ as
\begin{equation}
\begin{split}
&\langle (y(t_1)-\langle y(t_1)\rangle)(y(t_2)-\langle y(t_2)\rangle) \rangle\\
&~=\frac{F^2}{\Gamma(1+2\alpha)}t_1^{2\alpha}+\frac{2\sigma}{\Gamma(1+\alpha)}t_1^\alpha\\
&~~~~+\frac{F^2}{\Gamma^2(1+\alpha)} \left[_2F_1\left(\alpha, -\alpha; \alpha+1; \frac{t_1}{t_2}\right)-1\right]t_1^\alpha t_2^\alpha
\end{split}
\end{equation}
with a hypergeometric function  $_2F_1(a,b;c; z)$ \cite{AbramowitzStegun:1972}, which means that the process is non-stationary.  Using the MSD and autocorrelation function of process $y(t)$, one can obtain the ensemble averaged
integrand in \eqref{TAdefination2}
\begin{equation}
\begin{split}
&\langle ([y(t+\Delta)-y(t)]-\langle y(t+\Delta)-y(t)\rangle)^2 \rangle\\
=&\left(\frac{2F^2}{\Gamma(1+2\alpha)}-\frac{F^2}{\Gamma^2(1+\alpha)}\right)(t+\Delta)^{2\alpha}  \\
&+\frac{2\sigma}{\Gamma(1+\alpha)}(t+\Delta)^\alpha -\frac{F^2}{\Gamma^2(1+\alpha)}t^{2\alpha}\\
&-\frac{2\sigma}{\Gamma(1+\alpha)}t^\alpha+\frac{2F^2}{\Gamma^2(1+\alpha)}
\\
& \cdot\left[1-~_2F_1\left(\alpha, -\alpha; \alpha+1; \frac{t}{t+\Delta}\right)\right]t^\alpha(t+\Delta)^\alpha,
\end{split}
\end{equation}
which shows the aging phenomenon because of the explicit dependence on $t$.
%
Then we obtain the ensemble-time averaged MSD
\begin{equation}\label{TA_F}
\langle \overline{\delta^2(\Delta)}\rangle\simeq\frac{2\sigma}{\Gamma(1+\alpha)}\Delta T^{\alpha-1}+\frac{2F^2}{(1+\alpha)\Gamma^2(1+\alpha)}\Delta^{\alpha+1}T^{\alpha-1}
\end{equation}
for $\Delta \ll  T$. See the simulation results in Fig. \ref{ergo2_1}. For short lag time, the time averaged MSD is linearly dependent on lag time $\Delta$ as the free subdiffusion case in \eqref{TA-free-S}, which can be explained by the fact that the particles are not affected by external forces in short times. With the increase of lag time, the time averaged MSD becomes super-linear in lag time, and is proportional to the square of the external force.
The disagreement between time and ensemble averaged MSD shows the non-ergodicity of the stochastic process in \eqref{x_constant}, although it could exhibit ``superdiffusion'' phenomenon. Similar to the case of harmonic potential, the constant force in this model also only increases the coefficient of ergodicity breaking parameter,
\begin{equation}
\mathcal{EB}\simeq \frac{2\Gamma(2\alpha)}{(1+\alpha)(\alpha\Gamma(\alpha)^2-\Gamma(2\alpha))}\left(\frac{T}{\Delta}\right)^{\alpha-1}.
\end{equation}

For model \eqref{x_constant}, there is the generalized Einstein relation, being similar to the CTRW model for the subdiffusive process  \cite{HeBurovMetzlerBarkai:2008,FroembergBarkai:2013_nogo}, 
\begin{equation}\label{Einstein}
\langle \overline{\delta^1(\Delta)}\rangle_F=\frac{F}{2k_B \mathcal{T}}\langle\overline{\delta^2(\Delta)} \rangle_0,
\end{equation}
where $\langle\overline{\delta^1(\Delta)}\rangle_F=\frac{1}{T-\Delta}\int_0^{T-\Delta}\langle y(t+\Delta)-y(t)\rangle_F dt$.
From the discussions above, we recognize that if the generalized Einstein relation is satisfied by the original process $x(s)$ \cite{BouchaudGeorges:1990,BarkaiFleurov:1998}, it will still be valid for a subordinated process $y(t)$ in \eqref{x_constant}.

As for the correlation of process $y(t)$ in \eqref{x_constant},  by using the asymptotic expression $_2F_1(a, b; c; z)\simeq 1+\frac{ab}{c}z$ for small $z$, one could obtain that
\begin{equation}
\begin{split}
r[y(t_1),y(t_2)]\simeq G_2(t_1)t_2^{-\alpha},
\end{split}
\end{equation}
for fixed $t_1$ and $t_2\rightarrow \infty$. Here $G_2(t_1)$ could be regarded as a constant for fixed $t_1$. It indicates the long-range dependence of process $y(t)$, although the correlation is weaker than the one of free particle in \eqref{CC-free}.

\subsection{Force acting on subordinated process $y(t)$}
\begin{figure}
\begin{minipage}{0.35\linewidth}
  \centerline{\includegraphics[scale=0.248]{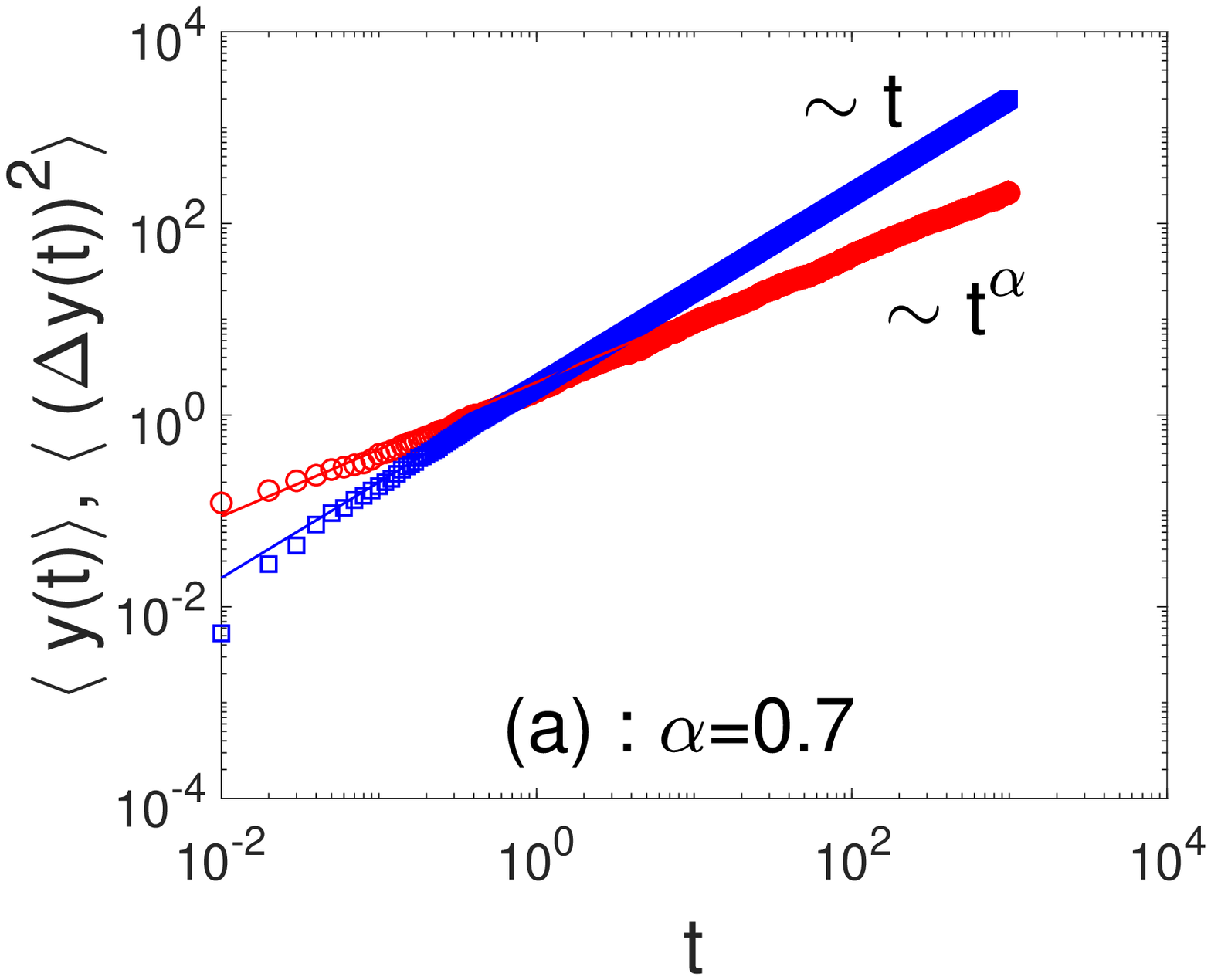}}
  \centerline{}
\end{minipage}
\hspace{1cm}
\begin{minipage}{0.35\linewidth}
  \centerline{\includegraphics[scale=0.248]{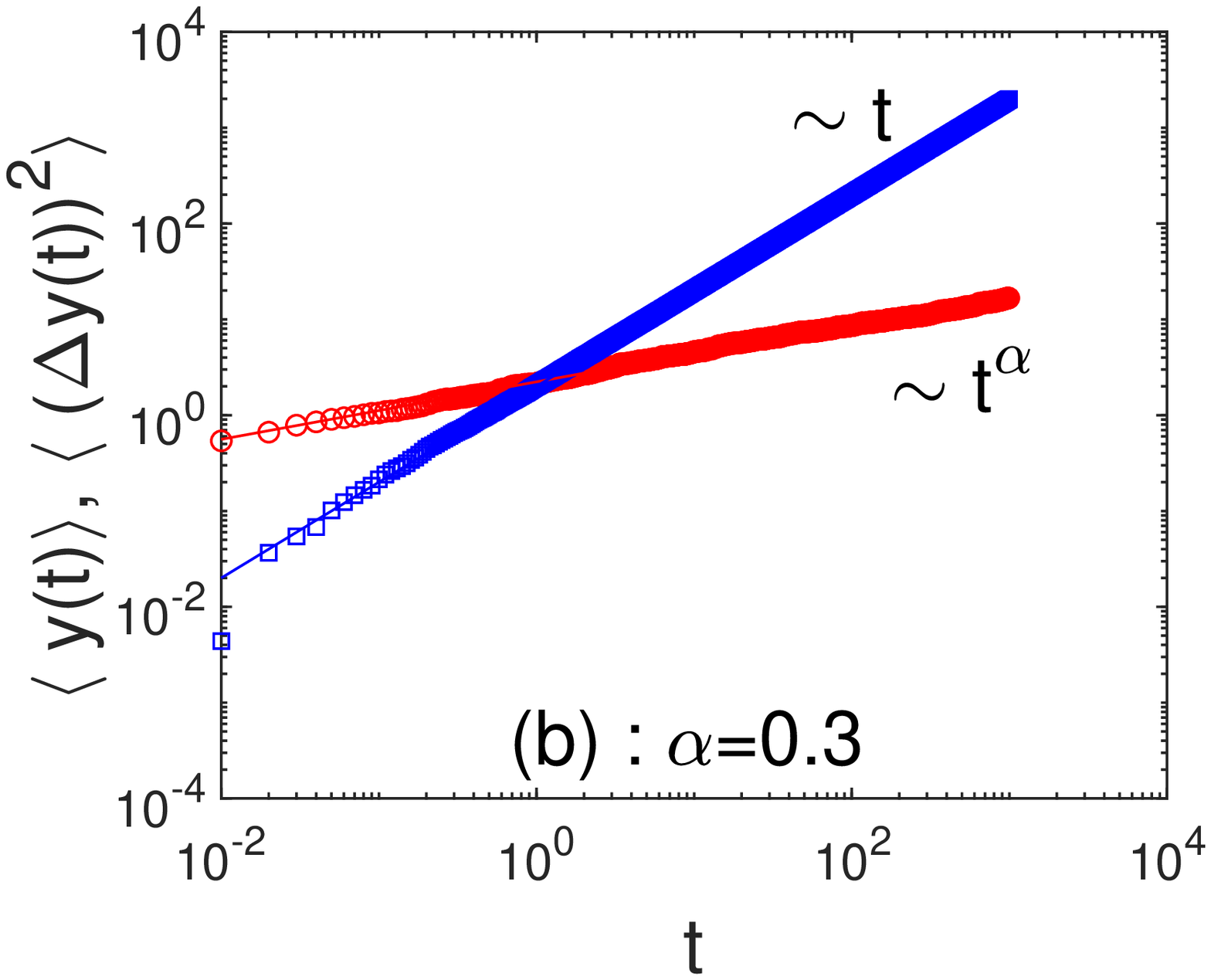}}
  \centerline{}
\end{minipage}
\caption{First moment and MSD of stochastic process described by the Langevin equation \eqref{y_constant2} for different $\alpha$. The parameters are, respectively, taken as $\sigma=1$, $F=2$, $\alpha=0.7$ (a) or $\alpha=0.3$ (b).
Blue solid lines and the blue square-markers represent the theoretical value $Ft$ and the simulation result of the first moment. Besides, the red solid lines and the red circle-markers represent the theoretical value \eqref{MSD_F2} and the simulation result of the ensemble averaged MSD.   }\label{MSD2_2}
\end{figure}

What about the differences if the external constant force $F$ acts directly on the subordinated process $y(t)$ and continues to affect the stochastic process all the time. In this case, the Langevin equation is \cite{CairoliBaule:2015_2,EuleFriedrich:2009}
\begin{equation}\label{y_constant2}
\dot{y}(t)=F+\sqrt{2\sigma}\overline{\xi}(t)
\end{equation}
with the equivalent coupled Langevin equation
\begin{equation}\label{x_constant2}
\dot{x}(s)=F \eta(s)+\sqrt{2\sigma}\xi(s), \qquad \dot{t}(s)=\eta(s).
\end{equation}
Using the solution of the exact trajectory, $y(t)=Ft+\sqrt{2\sigma}\int_0^t\overline{\xi}(\tau)d\tau$, the MSD of stochastic process $y(t)$ is
\begin{equation}\label{MSD_F2}
\langle(\Delta y(t))^2\rangle=\frac{2\sigma}{\Gamma(1+\alpha)}t^\alpha,
\end{equation}
which coincides with the MSD of free particle in \eqref{MSD-free}; see Fig. \ref{MSD2_2} for the simulation results. It implies that the external force does not change the subdiffusion behavior and behaves as a decoupled force. Hence, the subdiffusion model \eqref{y_constant2} is Galilei invariant \cite{MetzlerKlafter:2000}. In addition, the generalised Einstein relation in this case is not fulfilled since $\langle y(t)\rangle_F$ grows linearly with time $t$ while $\langle y^2(t)\rangle_0$ scales as $t^\alpha$.

The corresponding Fokker-Planck equation of model \eqref{y_constant2} is \cite{CairoliKlagesBaule:2018}
\begin{equation}\label{F}
\frac{\partial p(y,t)}{\partial t}=-F\frac{\partial p(y,t)}{\partial y}+\sigma \frac{\partial^2}{\partial y^2}\mathcal{D}_t^{1-\alpha}p(y,t),
\end{equation}
where
\begin{equation*}
\mathcal{D}_t^{1-\alpha}p(y,t)=\frac{1}{\Gamma(\alpha)}\left[\frac{\partial}{\partial t}+F\frac{\partial}{\partial y}\right]\int_0^t\frac{p(y-F(t-\tau), \tau)}{(t-\tau)^{1-\alpha}} d\tau
\end{equation*}
is fractional substantial derivative \cite{FriedrichJenkoBauleEulePRE:2006} with the Fourier-Laplace transform
\begin{equation*}
\mathcal{F}_{y\rightarrow k}[\mathcal{L}_{t\rightarrow \lambda}[\mathcal{D}_t^{1-\alpha}p(y,t)]]=(\lambda-ikF)^{1-\alpha}p(k, \lambda).
\end{equation*}
When $F=0$, $\mathcal{D}_t^{1-\alpha}$ recovers the Riemann-Liouville fractional derivative $D_t^{1-\alpha}$ and the Fokker-Planck equation goes back to the free subdiffusion case \eqref{FP_free}.

\begin{figure}
\begin{minipage}{0.35\linewidth}
  \centerline{\includegraphics[scale=0.248]{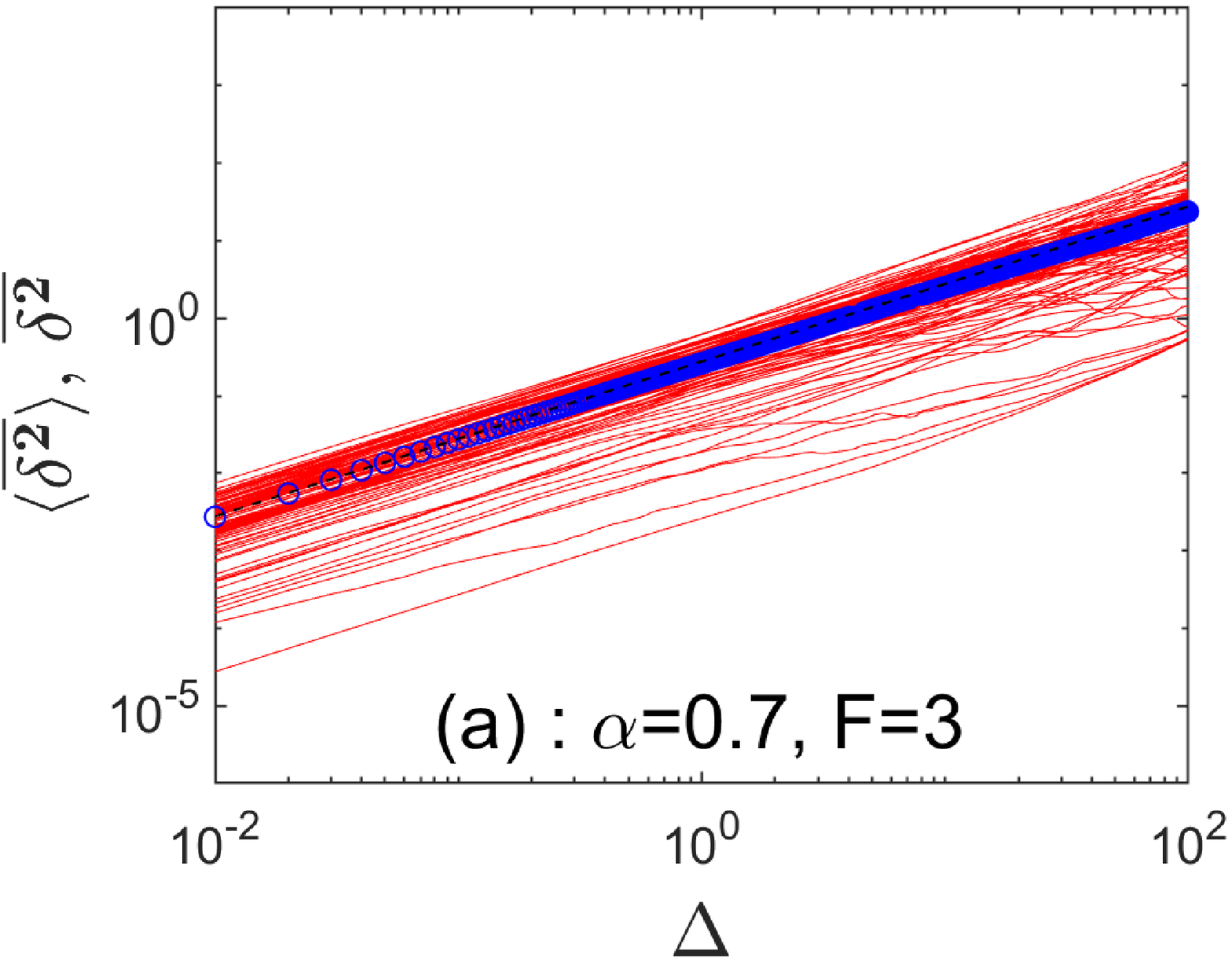}}
  \centerline{}
\end{minipage}
\hspace{1cm}
\begin{minipage}{0.35\linewidth}
  \centerline{\includegraphics[scale=0.248]{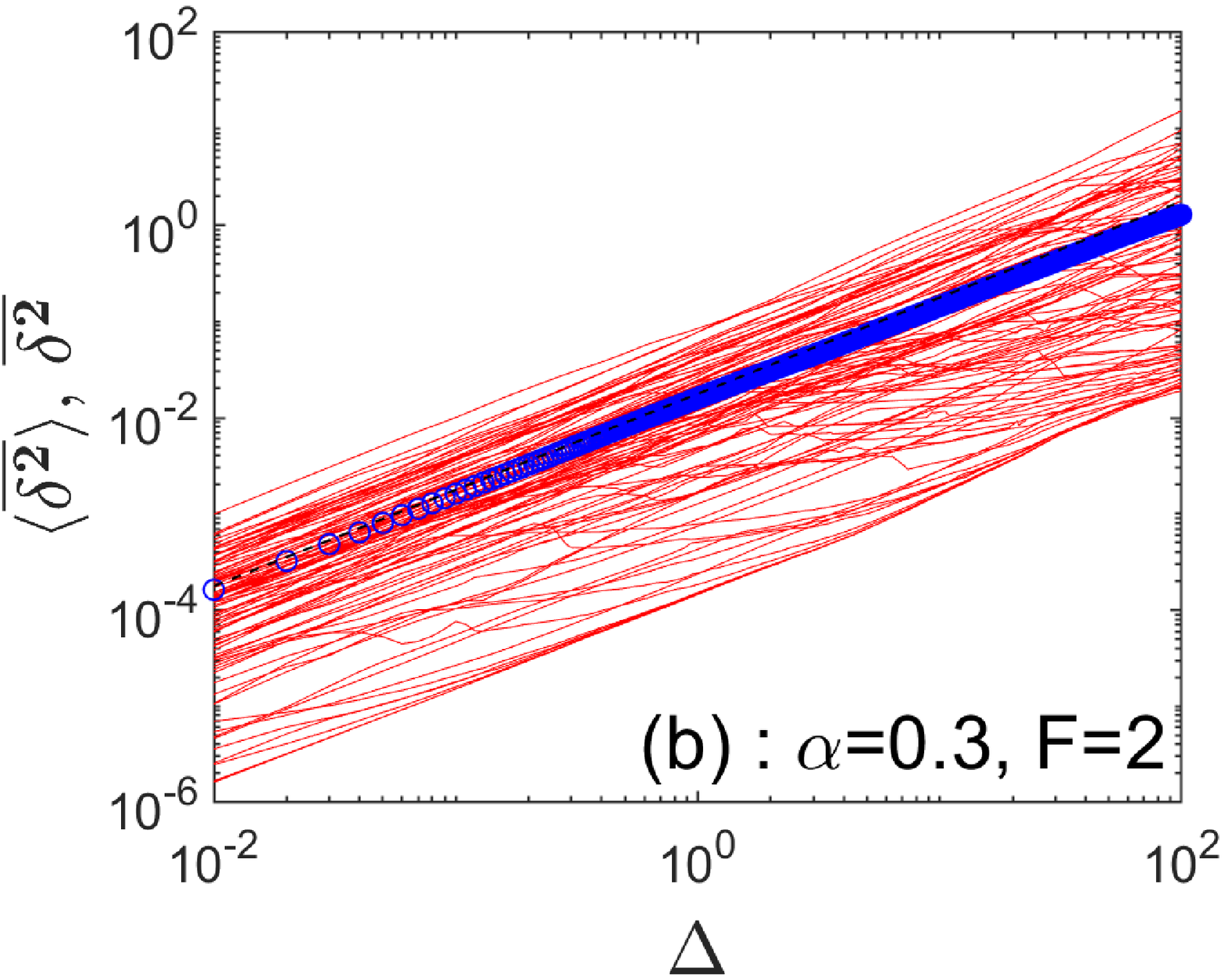}}
  \centerline{}
\end{minipage}
\caption{Time averaged MSD of the stochastic process described by Langevin equation \eqref{y_constant2}. Blue circle-markers represent the simulation results of the mean value of time averaged MSD and the red solid lines are the time averaged MSD of individual particle trajectories. Parameters are $T=1000$, $\sigma=1$, $\alpha=0.7$ and $F=3$ (a) or $\alpha=0.3$ and $F=2$ (b). The black dashed lines show the theoretical results \eqref{TA_F2}, which coincide with the simulation results of the ensemble-time averaged MSD over 100 trajectories. }\label{ergo2_2}
\end{figure}

After some simple calculations, we obtain the position autocorrelation function of the Langevin system \eqref{y_constant2}, $\langle (y(t_1)-\langle y(t_1)\rangle)(y(t_2)-\langle y(t_2)\rangle) \rangle=\frac{2\sigma}{\Gamma(1+\alpha)}t_1^\alpha$ for $t_2\geq t_1$, which reveals the non-stationary of this process. Then the correlation coefficient is obtained as
\begin{equation*}
r[y(t_1), y(t_2)]= \left(\frac{t_1}{t_2}\right)^{\alpha/2}
\end{equation*}
for fixed $t_1$ and large $t_2$, which is identical with the one of free particle in \eqref{CC-free}, implying that the decoupled force does not affect the correlation of positions at two different times.
In addition, the ensemble-time averaged MSD for $\Delta\ll T$ is
\begin{equation}\label{TA_F2}
\langle \overline{\delta^2(\Delta)}\rangle\simeq\frac{2\sigma}{\Gamma(1+\alpha)}\Delta T^{\alpha-1},
\end{equation}
which is the same as the free particle case in \eqref{TA-free-S}; see the simulation results in Fig. \ref{ergo2_2}. Obviously, the ergodicity breaking parameter also remains unchanged.
But the generalized Einstein relation is not satisfied when one compares $\langle \overline{\delta^1(\Delta)}\rangle_F =F\Delta$ with \eqref{TA_F2}.

The above results show that the decoupled constant force, which affects a stochastic process all the time,  will change the first moment of the stochastic process $y(t)$, but maintain the ensemble and time averaged MSD, as well as the correlation coefficient. The ergodic or non-ergodic behavior will not be changed in a decoupled force field. However, the Einstein relation is not valid any more due to the changes of the first moment.

Interestingly, the results for the case of the biased constant force in model \eqref{x_constant} are completely opposite. The ensemble and time averaged MSD are changed while the exponent of ergodicity breaking parameter is maintained. In addition, the Einstein relation is still valid.
On the other hand, the Fokker-Planck equation corresponding to the former model \eqref{x_constant} includes the Riemann-Liouville fractional derivative and the Fokker-Planck operator while the latter model \eqref{y_constant2} involves the fractional substantial derivative.

\section{Subdiffusive dynamics in time-dependent potential}\label{five}
This section focuses on the influence of the time-dependent periodic  oscillation force acting on the original process in \eqref{free1} with white Gaussian noise and on the subordinated process in \eqref{free2} with compound noise. Similar to another two kinds of forces discussed in Sec. \ref{three} and Sec. \ref{four}, the essential difference is still that the time-dependent force here only acts at the moments of jump for the former case, but keeps  acting on the system all the time for the latter case.

\subsection{Force acting on original process $x(s)$}
Consider the Langevin system with the time-dependent force acting on the original process $x(s)$, and it is expressed as \cite{MagdziarzWeronKlafter:2008,WeronMagdziarz:2008}
\begin{equation}\label{x_timedependent}
\dot{x}(s)=F(t(s))+\sqrt{2\sigma}\xi(s),\qquad \dot{t}(s)=\eta(s).
\end{equation}
Noting the subordination of the original process $x(s)$, the force term $F(t(s(t)))=F(t)$ has reasonable physical meaning, since a physical force should act on a system at physical time $t$ not internal time $s$ \cite{MagdziarzWeronKlafter:2008}. Its corresponding single Langevin equation describing the subordinated process $y(t)=x(s(t))$ in physical time is
\begin{equation}\label{y_timedependent}
\dot{y}(t)=F(t)\dot{s}(t)+\sqrt{2\sigma}\overline{\xi}(s).
\end{equation}
It is obvious that the time-dependent force $F(t)$ acts on the system only at the moments of jump, and the corresponding Fokker-Planck equation is \cite{SokolovKlafter:2006,MagdziarzWeronKlafter:2008,EuleFriedrich:2009,CairoliBaule:2017,SokolovKlafter:2006,Magdziarz:2009}
\begin{equation}\label{FP_timedependent}
\frac{\partial p(y,t)}{\partial t}=\left[-\frac{\partial}{\partial y}F(t)+\sigma \frac{\partial ^2}{\partial y^2}\right]D_t^{1-\alpha}p(y,t),
\end{equation}
where the Riemann-Liouville fractional derivative $D_t^{1-\alpha}$ cannot be interchanged with $-\frac{\partial}{\partial y}F(t)+\sigma \frac{\partial ^2}{\partial y^2}$.

Using the Fokker-Planck equation \eqref{FP_timedependent} derived from CTRW model, Sokolov \textit{et al.} \cite{SokolovKlafter:2006} obtained the recursive relation of the moments $r_n(t):=\langle y^n(t)\rangle$
\begin{equation}\label{recursive}
\frac{d r_n(t)}{dt}=nF(t)D_t^{1-\alpha}r_{n-1}(t)+\frac{n(n-1)}{2}D_t^{1-\alpha}r_{n-2}(t)
\end{equation}
with $r_0(t)=1, r_{-1}(t)=0,$ and $n\in\mathbb{N}$.
Two years later,  Magdziarz \textit{et al.} \cite{MagdziarzWeronKlafter:2008} derived the same recursive relation of the moments through the Langevin equation \eqref{y_timedependent}. Hence, the correspondence between the Fokker-Planck equation and the Langevin equation with a time-dependent force is established.

\begin{figure}
\begin{minipage}{0.5\linewidth}
  \centerline{\includegraphics[scale=0.32]{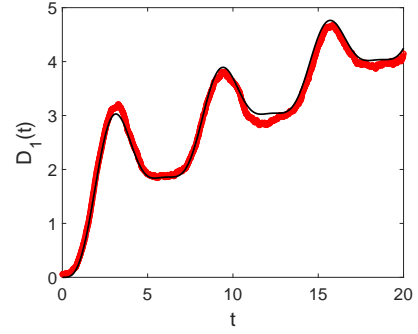}}
  \centerline{}
\end{minipage}
\caption{Fluctuation term of second moment of stochastic process described by the Langevin equation \eqref{x_timedependent}. Black solid line represents the analytical result $D_1(t)$ in \eqref{y2}, which coincides with the simulation result averaging over $10^4$ trajectories, represented by red circle-markers. Parameter values: $\alpha=0.7$, $\omega=1$, and $\sigma=1$.    }\label{MSD3_1}
\end{figure}

Here we take an oscillating external force $F(t)=f_0 \sin(\omega t)$. Although the ensemble averaged MSD of the stochastic process $y(t)$ can be obtained from the recursive relation \eqref{recursive}, here we also present the precise results of the position autocorrelation function and time averaged MSD by using the Laplace transform method.
Integrating \eqref{y_timedependent} with respect to time $t$ leads to
\begin{equation}
y(t)=\int_0^t F(t')ds(t')+\sqrt{2\sigma}B(s(t)).
\end{equation}
The position autocorrelation function of stochastic process $y(t)$ for $t_2 \geq t_1$ is
\begin{equation}\label{autocorrelation}
\begin{split}
&\langle y(t_1)y(t_2)\rangle  \\
&=\frac{2f_0^2}{\Gamma^2(\alpha)}\int_0^{t_1}\sin(\omega t_1')t_1'^{2\alpha-1}\\
&\cdot\int_0^1\sin(\omega t_1' u) u^{\alpha-1}(1-u)^{\alpha-1}dudt_1'\\
&+\frac{f_0^2}{\Gamma^2(\alpha)}\int_0^{t_1}\sin(\omega t_1')t_1'^{\alpha-1}\\
&\cdot\int_{t_1}^{t_2}(t_2'-t_1')^{\alpha-1}\sin(\omega t_2')dt_2'dt_1'+\frac{2\sigma}{\Gamma(1+\alpha)}t_1^\alpha,
\end{split}
\end{equation}
the detailed derivation of which is presented in Appendix \ref{App1}; and it shows the non-stationary property of the stochastic process $y(t)$. After long times, the asymptotic expression of position autocorrelation function can be obtained by taking $\lambda_1,\lambda_2\rightarrow0$ in \eqref{term1} and making inverse Laplace transform,
\begin{equation}
\langle y(t_1)y(t_2)\rangle\simeq\frac{f_0^2\cos(\frac{\pi}{2}\alpha)}{\Gamma(1+\alpha)\omega^\alpha}t_1^\alpha+\frac{2\sigma}{\Gamma(1+\alpha)}t_1^\alpha
\end{equation}
 for $t_2\geq t_1$.
Taking $t_1=t_2$ in \eqref{autocorrelation}, one obtains the second moment of $y(t)$,
\begin{equation}\label{y2}
\begin{split}
\langle &y^2(t)\rangle=\frac{2f_0^2}{\Gamma^2(\alpha)}\int_0^{t}\sin(\omega t')t'^{2\alpha-1}\\
&\cdot\int_0^1\sin(\omega t' u) u^{\alpha-1}(1-u)^{\alpha-1}dudt'
+\frac{2\sigma}{\Gamma(1+\alpha)}t^\alpha.
\end{split}
\end{equation}
In order to see the fluctuation of the second moment more clearly, we simulate the first term denoted as $D_1(t)$ in the second moment \eqref{y2}. The accordance between the simulation result and the analytical result could be found in Fig. \ref{MSD3_1}. For long times,
\begin{equation}\label{MSD-3}
\langle y^2(t)\rangle\simeq
\frac{f_0^2\cos(\frac{\pi}{2}\alpha)}{\Gamma(1+\alpha)\omega^\alpha}t^\alpha+\frac{2\sigma}{\Gamma(1+\alpha)}t^\alpha,
\end{equation}
which exhibits subdiffusion behavior, consistent with the asymptotic expression in \cite{SokolovKlafter:2006}.
The field-dependent contribution, which comes from the first term in \eqref{y2}, makes oscillation and additional dispersion of the particle position compared with the free subdiffusion case \eqref{MSD-free}. And for long times, this additional dispersion grows sublinearly with time $t$ in \eqref{MSD-3}.
In addition, the first moment of stochastic process $y(t)$ for long times can be easily obtained as
\begin{equation}\label{y1}
\begin{split}
\langle y(t)\rangle=\frac{f_0}{\Gamma(\alpha)}\int_0^t \sin(\omega t')t'^{\alpha-1}dt'
\simeq \frac{f_0}{\omega^\alpha}\sin\left(\frac{\pi}{2}\alpha\right).
\end{split}
\end{equation}
The oscillation of the mean value of process $y(t)$ tends to a constant for long times, which means the response to the external perturbation dies out for long times. It is also one of the manifestations of aging \cite{SokolovBlumenKlafter:2001}.

Another equivalent expression of stochastic process $y(t)$ in \eqref{y_timedependent} is
\begin{equation*}\label{y_t}
\dot{y}(t)=\int_0^\infty \delta(t-t(s)) F(t(s)) ds+\sqrt{2\sigma}\int_0^\infty \delta(t-t(s))\xi(s)ds.
\end{equation*}
Based on this expression, the position autocorrelation function of $y(t)$ can also be obtained. Detailed derivation is shown in Appendix \ref{App1}.

Combining the first two moments and the position autocorrelation function of the stochastic process $y(t)$, one obtains the correlation coefficient for fixed $t_1$ and $t_2\rightarrow \infty$
\begin{equation}
r[y(t_1), y(t_2)]\simeq \left(\frac{t_1}{t_2}\right)^{\alpha/2},
\end{equation}
which is the same as the free particle in \eqref{CC-free}. The essential reason for this interesting finding is that the  oscillating external force changes the coefficient of position autocorrelation function and ensemble averaged MSD of the free subdiffusive process at the same degree.

\begin{figure}
\begin{minipage}{0.35\linewidth}
  \centerline{\includegraphics[scale=0.248]{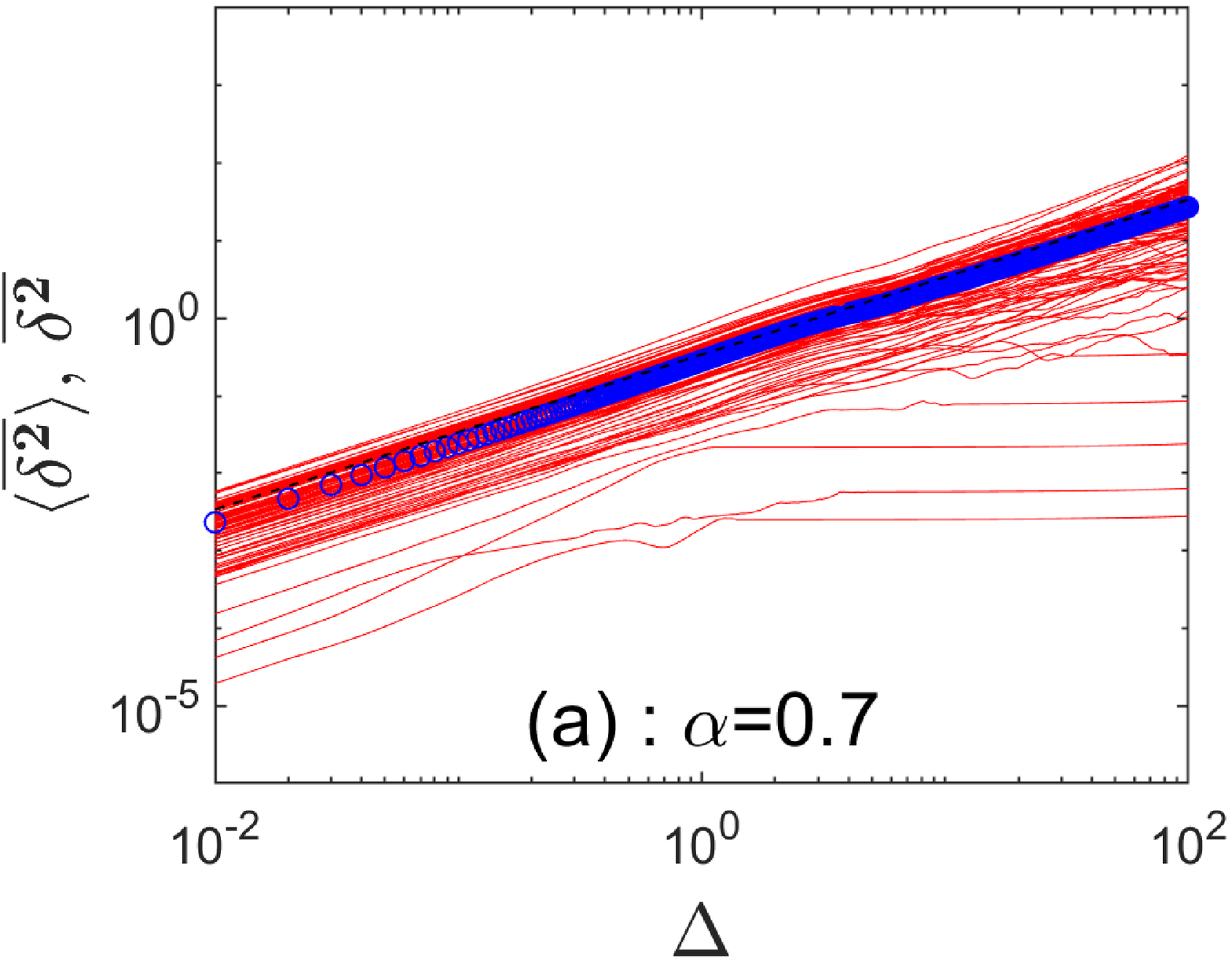}}
  \centerline{}
\end{minipage}
\hspace{1cm}
\begin{minipage}{0.35\linewidth}
  \centerline{\includegraphics[scale=0.248]{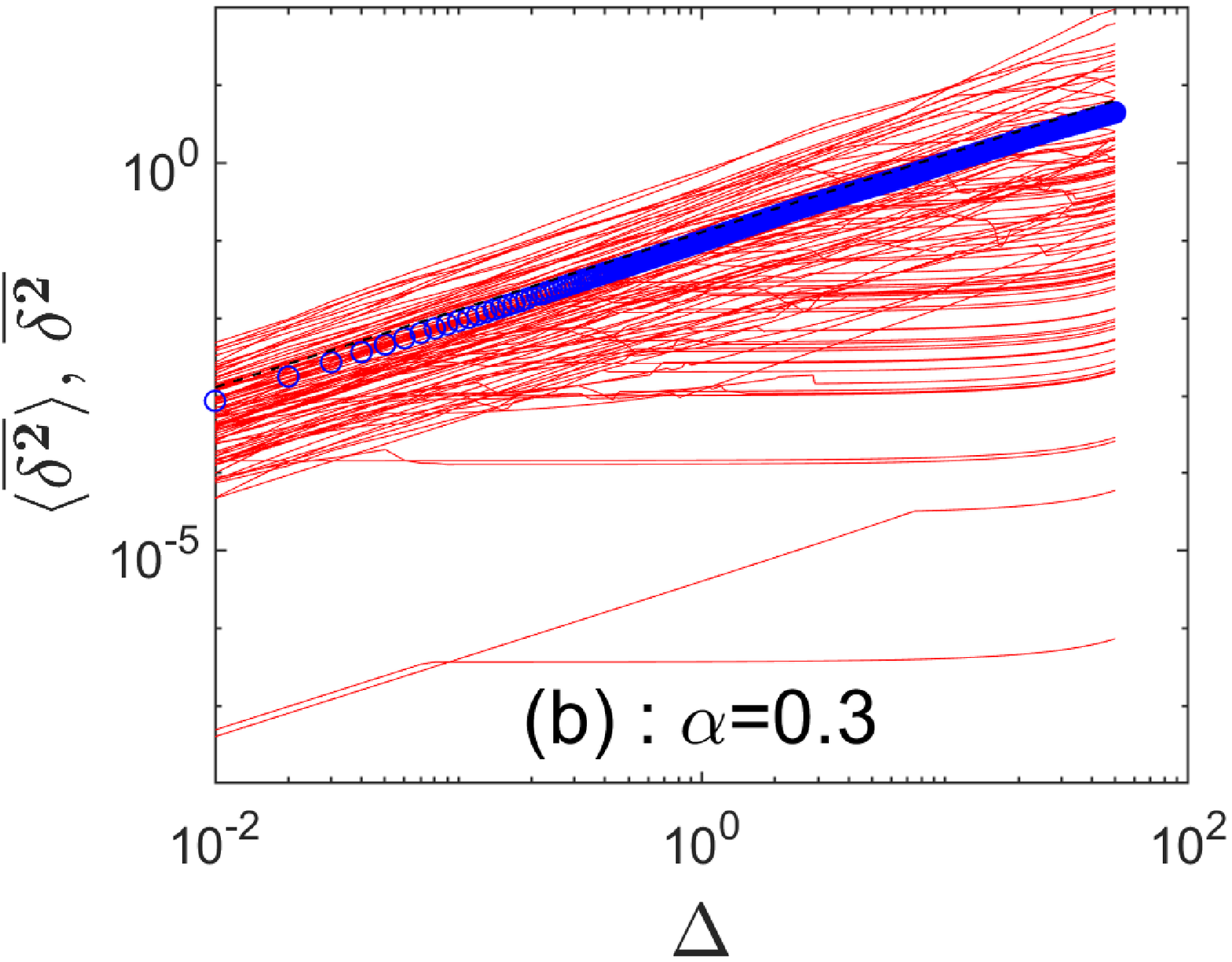}}
  \centerline{}
\end{minipage}
\caption{Simulation results of the time averaged MSD of stochastic process described by \eqref{x_timedependent} for different $\alpha$. The parameters are, respectively, taken as $\sigma=1$, $\omega=1$, $f_0=1$, $T=1000$, $\alpha=0.7$ (a) and $T=100$, $\alpha=0.3$ (b). The red solid lines represent the simulation results of time averaged MSD of individual trajectories and the blue circle-markers are the ensemble-time averaged MSD over 100 trajectories, which coincide with the theoretical results \eqref{TA_Ft} denoted by black dashed lines.
}\label{ergo3_1}
\end{figure}

The ensemble averaged integrand in \eqref{TAdefination2} for long times is
\begin{equation}
\begin{split}
&\langle ([y(t+\Delta)-y(t)]-\langle y(t+\Delta)-y(t)\rangle)^2 \rangle\\
&\simeq\frac{f_0^2\cos(\frac{\pi}{2}\alpha)+2\sigma\omega^\alpha}{\Gamma(1+\alpha)\omega^\alpha}[(t+\Delta)^\alpha-t^\alpha],
\end{split}
\end{equation}
the dependence of which on time $t$ implies the aging phenomenon of this Langevin system. The ensemble-time averaged MSD for $\Delta\ll T$ is
\begin{equation}\label{TA_Ft}
\langle \overline{\delta^2(\Delta)}\rangle \simeq \frac{f_0^2\cos(\frac{\pi}{2}\alpha)+2\sigma\omega^\alpha}{\Gamma(1+\alpha)\omega^\alpha}\Delta T^{\alpha-1}.
\end{equation}
See the simulation results in Fig. \ref{ergo3_1}.
Comparing with the case of free particle in \eqref{TA-free-S}, the oscillating external force here adds an additional contribution on the time averaged MSD, which grows linearly with the lag time $\Delta$. The disagreement between the time and ensemble averaged MSD, which scale as $\Delta$ and $\Delta^\alpha$ respectively, indicates the non-ergodicity behavior of the stochastic process. The ergodicity breaking parameter here is as same as the case of free particle due to the same degree of the changes on the coefficients of ensemble and time averaged MSD.

\subsection{Force acting on subordinated process $y(t)$}

Next, we consider the case in which the time-dependent external force acting on the system all the time, i.e.,
\begin{equation}\label{y_timedependent2}
\dot{y}(t)=F(t)+\sqrt{2\sigma}\overline{\xi}(t).
\end{equation}
The corresponding coupled Langevin equation is \cite{EuleFriedrich:2009}
\begin{equation}\label{x_timedependent2}
\dot{x}(s)=F(s)\eta(s)+\sqrt{2\sigma}\xi(s),\qquad \dot{t}(s)=\eta(s).
\end{equation}
In order to compare with the previous model \eqref{y_timedependent}, we also consider the oscillating force $F(t)=f_0\sin(\omega t)$ here.
The firstly moment of stochastic process $y(t)$ is
\begin{equation}\label{moment1}
\begin{split}
\langle y(t)\rangle=\frac{f_0}{w}(1-\cos(\omega t)),
\end{split}
\end{equation}
which shows a significant difference with the constant mean value \eqref{y1} of the model \eqref{y_timedependent}.
The mean value here keeps oscillation at a fixed frequency $\omega$ with the evolution of time since the oscillating external force $F(t)=f_0\sin(\omega t)$ influences this system for the whole time. The MSD of this model can be easily obtained as
\begin{equation}\label{moment2}
\langle(\Delta y(t))^2\rangle=\frac{2\sigma}{\Gamma(1+\alpha)}t^\alpha,
\end{equation}
which is identical with the case of free particle in \eqref{MSD-free}. It means that the time-dependent external force field here acts as a decoupled force, independent of the diffusion behavior. It is Galilean invariant model while another model \eqref{x_timedependent} breaks Galilean invariance.
The Fokker-Planck equation corresponding to the Langevin system \eqref{y_timedependent2} is
\begin{equation}\label{FP_At}
\begin{split}
\frac{\partial p(y,t)}{\partial t}=-\frac{\partial}{\partial y}F(t)p(y,t)+\sigma\frac{\partial^2}{\partial y^2}\mathcal{A}_t^{1-\alpha} p(y, t)
\end{split}
\end{equation}
with the operator in Fourier space
\begin{equation*}
\mathcal{F}_{y\rightarrow k}[\mathcal{A}_t^{1-\alpha} p(y, t)]=e^{ik \int_0^t F(t')dt'}D_t^{1-\alpha}e^{-ik \int_0^t F(t')dt'}p(k,t).
\end{equation*}
See the detailed derivations in Appendix \ref{App2}. Taking the constant force $F(t)=F$, the operator $\mathcal{A}_t^{1-\alpha}$ reduces to the fractional substantial derivative $\mathcal{D}_t^{1-\alpha}$ and the Fokker-Planck equation goes back to \eqref{F}.

\begin{figure}
\begin{minipage}{0.35\linewidth}
  \centerline{\includegraphics[scale=0.248]{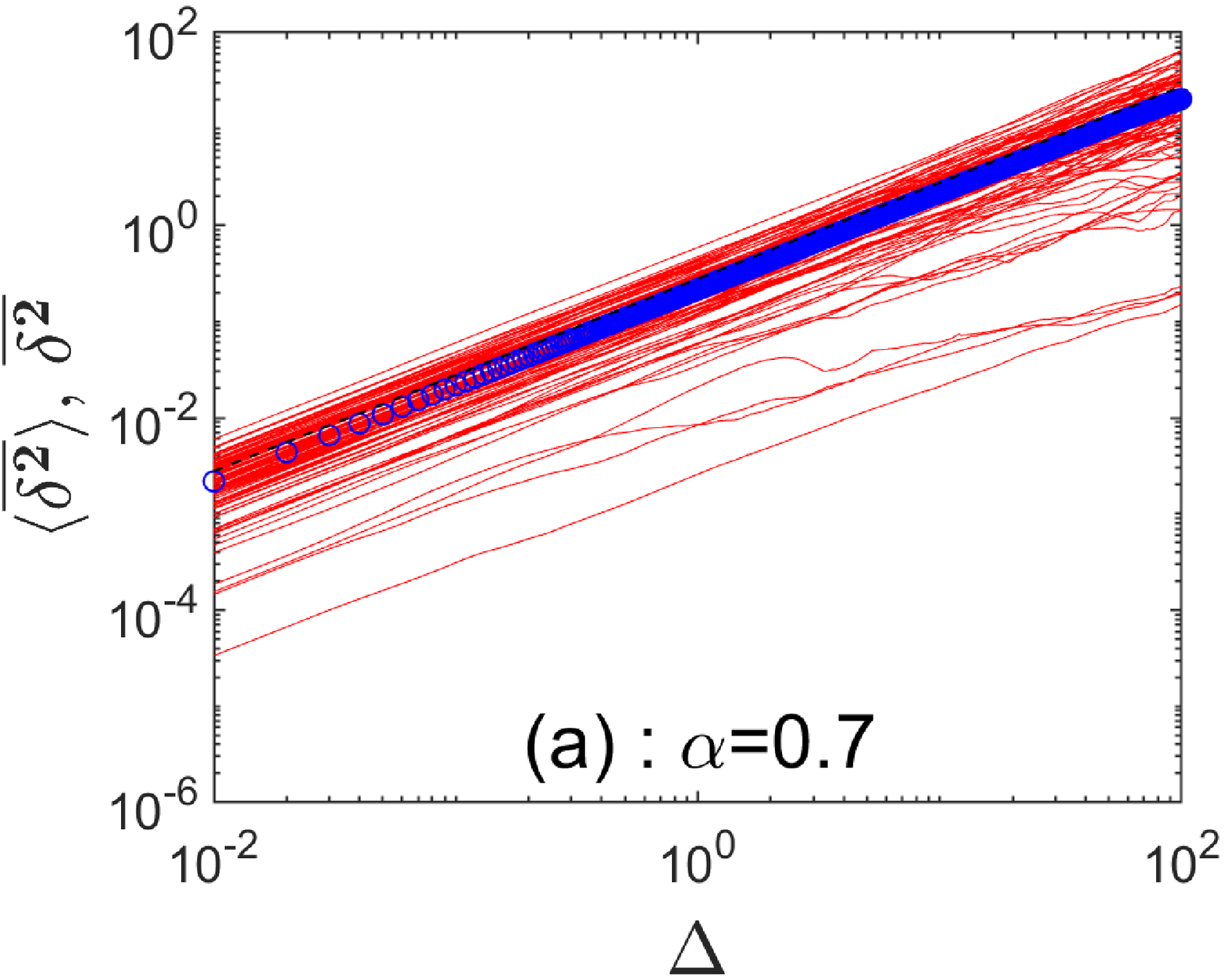}}
  \centerline{}
\end{minipage}
\hspace{1cm}
\begin{minipage}{0.35\linewidth}
  \centerline{\includegraphics[scale=0.248]{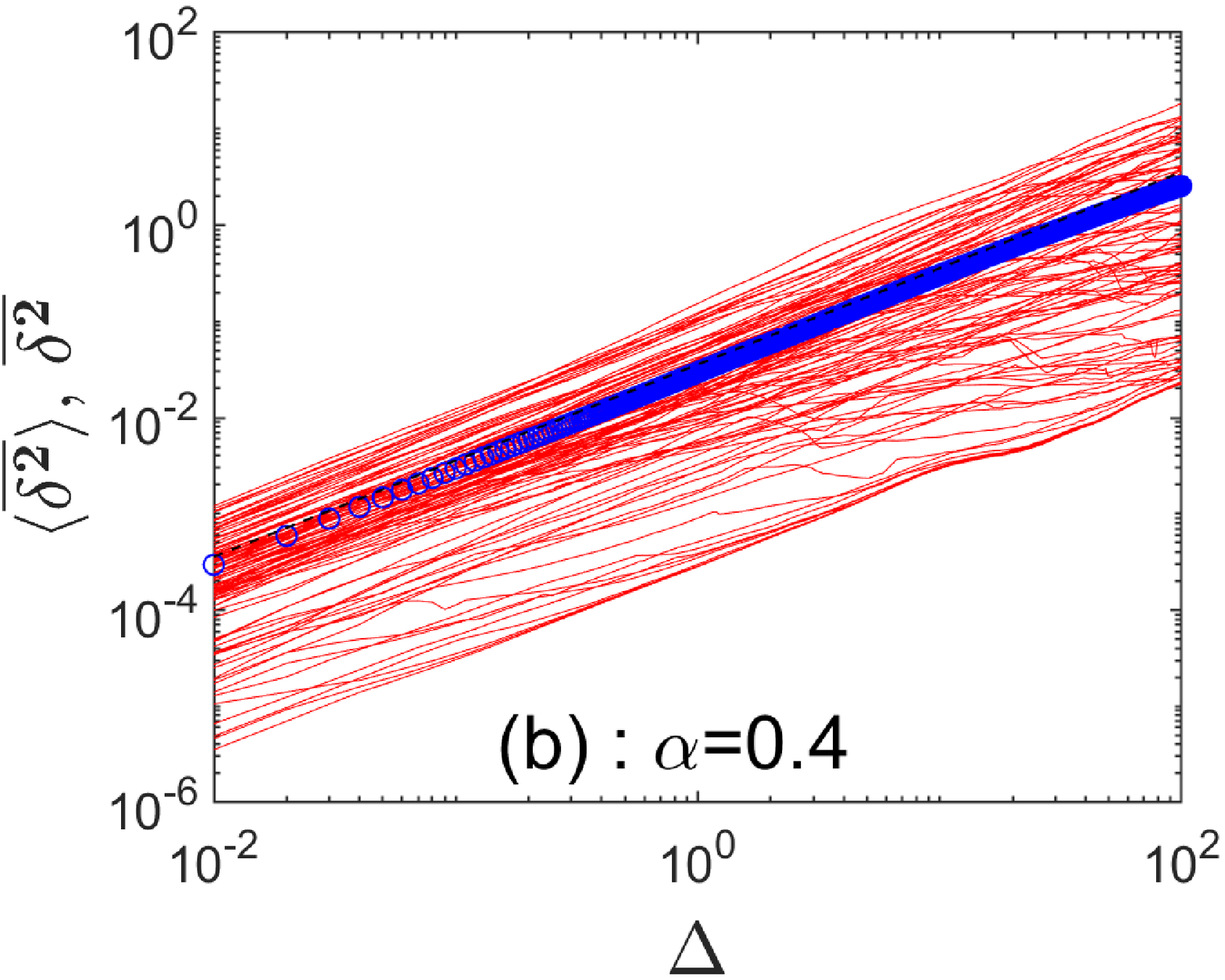}}
  \centerline{}
\end{minipage}
\caption{Simulation results of the time averaged MSD of stochastic process described by \eqref{y_timedependent2} for different $\alpha$. The parameters are, respectively, taken as $\sigma=1$, $\omega=1$, $f_0=1$, $T=1000$, $\alpha=0.7$ (a) and $\alpha=0.4$ (b). The red solid lines represent the simulation results of time averaged MSD of individual trajectories and the blue circle-markers are the ensemble-time averaged MSD over 100 trajectories, which coincide with the theoretical results represented by black dashed lines.
}\label{ergo3_2}
\end{figure}

As for the time averaged MSD, using the first two moments \eqref{moment1} and \eqref{moment2}, together with the position autocorrelation function
\begin{equation*}
\langle (y(t_1)-\langle y(t_1)\rangle)(y(t_2)-\langle y(t_2)\rangle)\rangle=\frac{2\sigma}{\Gamma(1+\alpha)}t_1^\alpha
\end{equation*}
for $t_2>t_1$, one could obtain the time averaged MSD for $\Delta\ll T$
\begin{equation*}
\langle \overline{\delta^2(\Delta)}\rangle \simeq \frac{2\sigma}{\Gamma(1+\alpha)}\Delta T^{\alpha-1},
\end{equation*}
which indicates the non-ergodic behavior of this Langevin system. It is also consistent with the case of free particle in \eqref{TA-free-S}. The simulation results could be found in Fig. \ref{ergo3_2}. In addition, the ergodicity breaking parameter and the correlation coefficient of this model are also unchanged, compared with the case of free particle.

In conclusion, the time-dependent force in \eqref{y_timedependent2} is a decoupled force, just as the decoupled constant force in \eqref{y_constant2}.
However, the time-dependent force in \eqref{x_timedependent} plays as a biasing force, just as the biased constant force in \eqref{x_constant}. They all change the ensemble and time averaged MSD of the free particle. One difference is that the time-dependent force keeps the same correlation coefficient as the free particle case, while the constant force weakens the correlation.
In addition, the Fokker-Planck equation corresponding to the former model \eqref{x_timedependent} includes the Riemann-Liouville fractional derivative while the latter model \eqref{y_timedependent2} involves a novel fractional derivative similar to the fractional substantial derivative.

\section{Summary}\label{six}

This paper focuses on the subdiffusion in an external force field. We mainly consider the influences of different patterns of external forces acting on the original process $x(s)$ or on the subordinated process $y(t)$. For this, we choose three kinds of common external forces --- linear force, constant force, and time-dependent oscillation force, and investigate some important statistical quantities depending on one-point or two-point PDF, such as ensemble and time averaged MSD, position autocorrelation function, correlation coefficient, and so on. There are obvious discrepancies between position-dependent and position-independent forces. 

One example of the position-dependent forces is the linear force (harmonic potential) in this paper. If it acts on original process $x(s)$, the ensemble averaged MSD tends to a non-zero constant for long times, while it tends to zero at power-law rate if this force acts on the subordinated process $y(t)$. The essential reason is that the external force drags the particle to zero position for all times in the latter case. These two stochastic processes are both non-ergodicity, non-stationary, and aging. However, the position at two different times in the former Langevin system is long-range dependent while in the latter it is not due to the continuous effects of external force. The position-dependent force acting on the original process $x(s)$ or on the process $y(t)$ does not affect the power of ergodicity breaking parameter, compared with the case of free particle.

As for the position-independent force, such as, constant force and periodic oscillation force in this paper, we find that it behaves as a biasing force if acting on the original process $x(s)$ and does change the ensemble and time averaged MSD, as well as the position autocorrelation function. One unexpected finding is the correlation coefficient --- it becomes weaker at the effects of constant force, while it remains unchanged in the case of the periodic oscillation force.
On the other hand, if the position-independent force acts on the subordinated process $y(t)$, it behaves as a decoupled force and does not make change of the statistical quantities we study.

Especially, the Fokker-Planck equations of the processes in different force field are different, being embodied by fractional derivatives; it is Riemann-Liouville type if the external force affects the process only at the moments of jump, but fractional substantial derivative or a similar novel fractional derivative if the external force keeps acting on the process all the time.

Collecting the properties of the statistical quantities and the discrepancies between different forces or different patterns the forces act on the system, we have a global knowledge of the motion of the subdiffusive particles in an external force field. This will help us to distinguish a large amount of processes with some similar features.

\section*{Acknowledgments}
This work was supported by the National Natural Science Foundation of China under grant no. 11671182, and the Fundamental Research Funds for the Central Universities under grant no. lzujbky-2018-ot03.

\appendix
\section{Derivation of the autocorrelation function \eqref{autocorrelation}}\label{App1}
The position autocorrelation function of process $y(t)$ is
\begin{equation}\label{Corr}
\begin{split}
  \langle y(t_1)y(t_2)\rangle
  &=\int_0^{t_1}\int_0^{t_2} F(t_1')F(t_2')\langle ds(t_1')ds(t_2')\rangle  \\
  &~~~+\langle B(s(t_1))B(s(t_2))\rangle,
\end{split}
\end{equation}
where the cross-terms are zero due to the independence of $B(s)$ and $s(t)$ and could be omitted. For the first term in \eqref{Corr}, denoted as $\langle y(t_1)y(t_2)\rangle_1$, it can be dealt with in Laplace space. Using the expression of
\begin{equation}\label{Ft1Ft2}
\begin{split}
F(t_1)F(t_2)&=f_0^2\sin(\omega t_1)\sin(\omega t_2)\\
&=-\frac{f_0^2}{4}(e^{i \omega t_1}e^{i \omega t_2}-e^{i \omega t_1}e^{-i \omega t_2}\\
&~~~~-e^{-i \omega t_1}e^{i \omega t_2}+e^{-i \omega t_1}e^{-i \omega t_2}),
\end{split}
\end{equation}
and the correlation function of inverse subordinator $s(t)$ in Laplace space \cite{BauleFriedrich:2005}
\begin{equation}
\begin{split}
&\mathcal{L}_{t_1\rightarrow \lambda_1, t_2\rightarrow\lambda_2}[\langle s(t_1)s(t_2)\rangle]   \\
&~~~~~=(\lambda_1+\lambda_2)^{-\alpha}\left( \frac{\lambda_1^{-\alpha-1}}{\lambda_2}+\frac{\lambda_2^{-\alpha-1}}{\lambda_1}\right),
\end{split}
\end{equation}
one get
\begin{equation}\label{term1}
\begin{split}
&\mathcal{L}_{t_1\rightarrow\lambda_1, t_2\rightarrow\lambda_2}[\langle y(t_1)y(t_2)\rangle_1]\\
&=-\frac{f_0^2}{4}\frac{1}{\lambda_1\lambda_2}\left[ \frac{1}{(\lambda_1^-+\lambda_2^-)^\alpha(\lambda_1^-)^\alpha}+\frac{1}{(\lambda_1^-+\lambda_2^-)^\alpha(\lambda_2^-)^\alpha}  \right.\\
&~~~-\frac{1}{(\lambda_1+\lambda_2)^\alpha(\lambda_1^-)^\alpha}-\frac{1}{(\lambda_1+\lambda_2)^\alpha(\lambda_2^+)^\alpha}\\
&~~~-\frac{1}{(\lambda_1+\lambda_2)^\alpha(\lambda_1^+)^\alpha}-\frac{1}{(\lambda_1+\lambda_2)^\alpha(\lambda_2^-)^\alpha}\\
&~~~+\left.\frac{1}{(\lambda_1^++\lambda_2^+)^\alpha(\lambda_1^+)^\alpha}+\frac{1}{(\lambda_1^++\lambda_2^+)^\alpha(\lambda_2^+)^\alpha}\right],
\end{split}
\end{equation}
where $\lambda_j^\pm=\lambda_j\pm iw$ with $j=1,2$.
After performing the inverse Laplace transform, the first term of  position autocorrelation function is
\begin{equation}\label{autocorrelation1}
\begin{split}
&\langle y(t_1)y(t_2)\rangle_1\\
&=\frac{2f_0^2}{\Gamma^2(\alpha)}\int_0^{t_1}\sin(\omega t_1')t_1'^{2\alpha-1}\\
&\cdot\int_0^1\sin(\omega t_1' u) u^{\alpha-1}(1-u)^{\alpha-1}dudt_1'\\
&+\frac{f_0^2}{\Gamma^2(\alpha)}\int_0^{t_1}\sin(\omega t_1')t_1'^{\alpha-1}\\
&\cdot\int_{t_1}^{t_2}(t_2'-t_1')^{\alpha-1}\sin(\omega t_2')dt_2'dt_1',
\end{split}
\end{equation}
where
\begin{equation*}
\begin{split}
\int_0^1&\sin(\omega t_1' u) u^{\alpha-1}(1-u)^{\alpha-1}du
=\frac{\sqrt{\pi} t' \omega \Gamma(\alpha)}{\Gamma(\frac{1}{2}+\alpha)2^{2\alpha}}\\
&\cdot {}_2F_3\left(\frac{1+\alpha}{2},\frac{2+\alpha}{2}; \frac{3}{2}, \frac{1}{2}+\alpha, 1+\alpha; -\frac{1}{4}t'^2\omega^2\right)
\end{split}
\end{equation*}
for $t_2 \geq t_1$.
The second term of $\langle y(t_1)y(t_2)\rangle$ is
\begin{equation*}
\begin{split}
\langle y(t_1)y(t_2)\rangle_2
&=2\sigma\langle B(s(t_1))B(s(t_2))\rangle\\
&=2\sigma\min\{\langle s(t_1)\rangle,\langle s(t_2)\rangle\}\\
&=\frac{2\sigma}{\Gamma(1+\alpha)}\min\{t_1^\alpha, t_2^\alpha\}.
\end{split}
\end{equation*}
Here we use the independence of $B(s)$ and $s(t)$, as well as $\langle s(t)\rangle=\frac{1}{\Gamma(1+\alpha)}t^\alpha$ \cite{BauleFriedrich:2005}.
Finally, the position autocorrelation function of stochastic process $y(t)$ for $t_2 \geq t_1$ is \eqref{autocorrelation}.

We now present another exact expression of $y(t)=x(s(t))$. Similar to the method in \cite{CairoliBaule:2015_2},
\begin{equation}\label{y}
\begin{split}
y(t)=&\int_0^{s(t)} \dot{x}(\tau)d\tau\\
=&\int_0^\infty\delta(s-s(t))\int_0^s F(t(\tau))d\tau ds\\
&+\sqrt{2\sigma}\int_0^\infty\delta(s-s(t))\int_0^s \xi(\tau)d\tau ds\\
=&\int_0^\infty \Theta(t-t(s)) F(t(s)) ds\\
&+\sqrt{2\sigma}\int_0^\infty \Theta(t-t(s))\xi(s)ds,
\end{split}
\end{equation}
where the last equality is obtained by using $\delta(s-s(t))=-\frac{\partial}{\partial s}\Theta(t-t(s))$ \cite{BauleFriedrich:2005} and integration by parts.
Then the differential expression of process $y(t)$ is
\begin{equation}\label{ydot}
\begin{split}
\dot{y}(t):&=\dot{y}_1(t)+\dot{y}_2(t)\\
&=\int_0^\infty \delta(t-t(s)) F(t(s)) ds  \\
&~~~+\sqrt{2\sigma}\int_0^\infty \delta(t-t(s))\xi(s)ds.
\end{split}
\end{equation}
The first term of the autocorrelation function of $y(t)$ \eqref{term1} in Laplace space could also be obtained from the first term in \eqref{ydot}:
\begin{equation*}
\begin{split}
&\langle \dot{y}_1(t_1)\dot{y}_1(t_2)\rangle\\
&=\int_0^\infty \int_0^\infty \int_0^\infty \int_0^\infty \delta(t_1-t(s_1))\delta(t_2-t(s_2)) \\
&~\cdot F(t(s_1))F(t(s_2)) p(t(s_1), t(s_2), s_1, s_2)dt(s_1)dt(s_2)ds_1ds_2\\
&=F(t_1)F(t_2)\int_0^\infty \int_0^\infty p(t_1, t_2, s_1, s_2)ds_1ds_2,
\end{split}
\end{equation*}
where $p(t(s_1), t(s_2), s_1, s_2)$ is the two-point joint PDF of subordinator $t(s)$. Using the expression of $p(t_1, t_2, s_1, s_2)$ in Laplace space \cite{BauleFriedrich:2005}, i.e.,
\begin{equation*}
\begin{split}
p(\lambda_1, \lambda_2, s_1, s_2)&=\Theta(s_2-s_1)e^{-s_1(\lambda_1+\lambda_2)^\alpha}e^{-(s_2-s_1)\lambda_2^\alpha}\\
&+\Theta(s_1-s_2)e^{-s_2(\lambda_1+\lambda_2)^\alpha}e^{-(s_1-s_2)\lambda_1^\alpha},
\end{split}
\end{equation*}
and \eqref{Ft1Ft2}, we can also obtain the term \eqref{term1} of autocorrelation function of $y(t)$ in Laplace space.

\section{Derivation of the Fokker-Planck equation \eqref{FP_At}}\label{App2}
By the similar method shown in \cite{CairoliBaule:2017,CairoliBaule:2015}, we now derive the Fokker-Planck equation corresponding to the Langevin equation $\dot{y}(t)=F(t)+\sqrt{2\sigma}\overline{\xi}(t)$, equivalently, $y(t)=\int_0^t F(t')dt'+\sqrt{2\sigma}B(s(t))$. As we all know, $\int_0^t F(t')dt'$ is a process with finite variation and $B(s(t))$ is a martingale, which lead process $y(t)$ to be a semi-martingale \cite{Kunita:1997}. In addition, $y(t)$ has continuous path. Hence we can use its It\^{o} formula as follows \cite{Kunita:1997}
\begin{equation}
\begin{split}
f(y(t))&=f(y_0)+\int_0^t f'(y(\tau))dy(\tau)   \\
&~~~+\frac{1}{2}\int_0^t f''(y(\tau))d[y, y]_\tau,
\end{split}
\end{equation}
where $[y, y]_t=\sum_i|y(t_i)-y(t_{i-1})|^2=2\sigma\int_0^tds(\tau)$ is the quadratic variation of process $y(t)$ \cite{CairoliBaule:2017,Oksendal:2005}, and it could be gotten by $dtdt=dtdB_t=dB_tdt=0$.
The PDF of process $y(t)$ in Fourier space is $p(k, t)=\mathcal{F}_{y\rightarrow k}[\langle\delta(y-y(t))\rangle]=\langle e^{iky(t)}\rangle$. So we take $f(y(t))=e^{iky(t)}$. Then
\begin{equation}\label{Ito}
\begin{split}
e^{iky(t)}&=e^{iky_0}+ik\int_0^t e^{iky(\tau)}dy(\tau)-\sigma k^2\int_0^t e^{iky(\tau)}ds(\tau)\\
&=e^{iky_0}+ik\int_0^t e^{iky(\tau)}F(\tau)d\tau      \\
&~~~ + ik\sqrt{2\sigma}\int_0^t e^{iky(\tau)}\overline{\xi}(\tau)d\tau - \sigma k^2\int_0^t e^{iky(\tau)}ds(\tau).
\end{split}
\end{equation}
Taking the ensemble average of \eqref{Ito}, making inverse Fourier transform, and taking partial derivative with respect to $t$, we obtain 
\begin{equation}\label{FPequation}
\begin{split}
\frac{\partial p(y,t)}{\partial t}=-\frac{\partial}{\partial y}F(t)p(y,t)+\sigma\frac{\partial^2}{\partial y^2}\langle \delta(y-y(t))\dot{s}(t)\rangle.
\end{split}
\end{equation}
The last term $\langle \delta(y-y(t))\dot{s}(t)\rangle$ could be dealt with in Fourier space
\begin{equation}
\begin{split}
&\langle e^{iky(t)}\dot{s}(t)\rangle=e^{ik \int_0^t F(t')dt'} \left\langle e^{ik \sqrt{2\sigma}B(s(t))}\dot{s}(t)\right\rangle\\
&=e^{ik \int_0^t F(t')dt'} \\
&~~\cdot\frac{\partial}{\partial t}\left\langle \int_0^t\left(\int_0^\infty  e^{ik \sqrt{2\sigma}B(s)} \delta(s-s(\tau))ds\right)ds(\tau)\right\rangle\\
&=e^{ik \int_0^t F(t')dt'}\frac{\partial}{\partial t}\int_0^\infty\langle e^{ik \sqrt{2\sigma}B(s)}\rangle\langle \Theta(t-t(s))\rangle ds\\
&=e^{ik \int_0^t F(t')dt'} \int_0^\infty\langle e^{ik \sqrt{2\sigma}B(s)}\rangle\langle \delta(t-t(s))\rangle ds.\\
\end{split}
\end{equation}
Here we use the fact $\int_0^t   \delta(s-s(\tau))ds(\tau)=\Theta(t-t(s))$ since $\Theta(t-t(s))=1-\Theta(s-s(t))$ \cite{BauleFriedrich:2005,BauleFriedrich:2007}.
For simplicity of notation, let us define $G_1(k, t)=\int_0^\infty\langle e^{ik \sqrt{2\sigma}B(s)}\rangle\langle \delta(t-t(s))\rangle ds$. Taking Laplace transform, one has
\begin{equation}\label{G1}
\begin{split}
G_1(k, \lambda)=\int_0^\infty\langle e^{ik \sqrt{2\sigma}B(s)}\rangle e^{-s\lambda^\alpha} ds.
\end{split}
\end{equation}
Similarly, the PDF of $y(t)$ in Fourier space is
\begin{equation}
\begin{split}
\langle e^{iky(t)}\rangle=e^{ik \int_0^t F(t')dt'} \int_0^\infty\langle e^{ik \sqrt{2\sigma}B(s)}\rangle\langle \delta(s-s(t))\rangle ds.
\end{split}
\end{equation}
We define $G_2(k, t)=\int_0^\infty\langle e^{ik \sqrt{2\sigma}B(s)}\rangle\langle \delta(s-s(t))\rangle ds$, the Laplace transform of which is
\begin{equation}
\begin{split}
G_2(k, \lambda)
=\lambda^{\alpha-1}\int_0^\infty\langle e^{ik \sqrt{2\sigma}B(s)}\rangle e^{-s\lambda^\alpha} ds.
\end{split}
\end{equation}
Here we use the PDF of inverse subordinator $s(t)$ in Laplace space \eqref{hslambda}.
Therefore, $G_1(k,\lambda)=\lambda^{1-\alpha}G_2(k,\lambda)$, i.e.,
\begin{equation}
G_1(k,t)=D_t^{1-\alpha}G_2(k,t).
\end{equation}
 Then one has
\begin{equation}
\begin{split}
\langle e^{iky(t)}\dot{s}(t)\rangle=e^{ik \int_0^t F(t')dt'}D_t^{1-\alpha}e^{-ik \int_0^t F(t')dt'}\langle e^{iky(t)}\rangle.
\end{split}
\end{equation}
Finally, the  Fokker-Planck equation corresponding to the Langevin system $\dot{y}(t)=F(t)+\sqrt{2\sigma}\overline{\xi}(t)$ is
\begin{equation}
\begin{split}
\frac{\partial p(y,t)}{\partial t}=-\frac{\partial}{\partial y}F(t)p(y,t)+\sigma\frac{\partial^2}{\partial y^2}\mathcal{A}_t p(y, t),
\end{split}
\end{equation}
where the operator $\mathcal{A}_t$ in Fourier space is  $\mathcal{F}_{y\rightarrow k}[\mathcal{A}_t p(y, t)]=e^{ik \int_0^t F(t')dt'}D_t^{1-\alpha}e^{-ik \int_0^t F(t')dt'}p(k,t)$.

\section*{References}
\bibliographystyle{apsrev4-1}
\bibliography{Reference}

\end{document}